\documentclass[useAMS,usenatbib]{mn2e}

\usepackage{graphicx}
\usepackage{epsfig} 
\usepackage{amssymb}
\usepackage{textcomp}

\title[A cellular automaton model of pulsar glitches]{A cellular automaton model of pulsar glitches}
\author[L. Warszawski and A. Melatos]{L. Warszawski$^{1}$\thanks{lilaw@ph.unimelb.edu.au (LW)} and A. Melatos$^{1}$\thanks{amelatos@unimelb.edu.au (AM)}\\
$^{1}$School of Physics, University of Melbourne, Parkville, VIC 3010, Australia\\}
\begin{document}

\date{}

\pagerange{\pageref{firstpage}--\pageref{lastpage}} \pubyear{2007}

\maketitle

\label{firstpage}

\begin{abstract}
A cellular automaton model of pulsar glitches is described, based on the superfluid vortex unpinning paradigm.  Recent analyses of pulsar glitch data suggest that glitches result from scale-invariant avalanches \citep{Melatos07a}, which are consistent with a self-organized critical system (SOCS).  A cellular automaton provides a computationally efficient means of modelling the collective behaviour of up to $10^{16}$ vortices in the pulsar interior, whilst ensuring that the dominant aspects of the microphysics are not lost.  The automaton generates avalanche distributions that are qualitatively consistent with a SOCS and with glitch data.  The probability density functions of glitch sizes and durations are power laws, and the probability density function of waiting times between successive glitches is Poissonian, consistent with statistically independent events.  The output of the model depends on the physical and computational paramters used.  The fitted power law exponents for the glitch sizes ($a$) and durations ($b$) decreases as the strength of the vortex pinning increases.  Similarly the exponents increase as the fraction of vortices that are pinned decreases.  For the physical and computational parameters considered, one finds $-4.3\leq a \leq -2.0$ and $-5.5\leq b\leq -2.2$, and mean glitching rates in the range $0.0023\leq\lambda\leq0.13\rm$ in units of inverse time.  \newline\newline
\end{abstract}


\section{Introduction}
\label{sec:intro}

It has recently been shown that glitches in individual pulsars obey avalanche statistics \citep{Melatos07a}.   Such statistical behaviour, in conjunction with the wide dynamical range of glitches, suggests that the underlying physics is that of a self-organized critical system (SOCS) \citep{Pines85,Bak96,Jensen98,Melatos07a}.  SOCS are characterised by transitions between metastable states via scale invariant avalanches \citep{Jensen98}.  The fractional increase in spin frequency spans seven decades across the entire glitch population ($10^{-11}\leq\delta\nu / \nu\leq 10^{-4}$), and up to four decades in any individual pulsar.  A natural way to model a scale invariant, avalanching system is to construct a cellular automaton that is driven slowly to a threshold of instability \citep{Field}.  Such models have been explored in detail in the context of earthquakes \citep{Sornette}, granular assemblies \citep{Bak88,Wiesenfeld,Bak96,Frette96,Pruessner03}, solar flares \citep{Lu}, and superconducting vortices \citep{Jensen,Field,Bassler,Linder}.

Theories of pulsar glitches centre around the mass unpinning model first proposed by \cite{Anderson} and extended by many others \citep{Alpar84,Pines85,Alpar89,Link02}.  The neutron superfluid in the stellar interior is threaded by many ($\sim 10^{16}$) vortices, approximately one percent of which are pinned to the stellar crust at grain boundaries and/or nuclear lattice sites \citep{Alpar89}.  As the pulsar crust spins down electromagnetically, a lag builds up between the velocity of the pinned vortex lines (corotating with the crust) and the superfluid.  When the transverse Magnus force (directly proportional to the lag) surpasses a threshold value (equal to the strength of the pinning force), a catastrophic unpinning of vortices occurs, transferring angular momentum to the crust.  In order for this mechanism to generate glitches on the scale observed, it requires up to $10^{12}$ vortices to unpin simultaneously, exhibiting a high level of collective, non-local behaviour.

The mass unpinning model spawned much activity in quantifying the microphysics that would lead to such macroscopic behaviour \citep{Pines85,Jones97,Jones98a,Jones98b,DeB99,Elgaroy01,Donati03,Avogadro}.  In particular, the strength of the pinning force has recently been discussed in detail using a self-consistent, semi-analytical model for the vortex-nucleus interaction \citep{Avogadro}, yielding the unexpected result that the strongest pinning occurs in regions of the crust that are of high and low density, rather than intermediate density \citep{Donati06}.  An interest has also been taken in the evolution and morphology of the neutron star crust, with particular attention paid to impurities and defects, which determine the density of pinning centres available to the vortices, and their relative strengths \citep{DeB98}.  \cite{Link02} looked at the role that precession, if present, plays in the unpinning process, and discussed the dependence of the pinnning force on various microphysical parameters of the superfluid vortices and of the nuclear crustal lattice.  

Several other theories of pulsar glitches, similarly focused on the microphysics of the superfluid in the stellar interior and crust, are currently under consideration.  These include thermally driven glitches arising from sudden changes in the frictional coupling within the neutron star \citep{Link96}, superfluid turbulence \citep{Peralta05,Peralta06a,Peralta06b,Melatos07b}, superfluid two-stream instabilities \citep{Andersson04,Mastrano05}, crust cracking \citep{Ruderman69,Bak96,Alpar96,Middleditch06} and mass accretion \citep{Morley96a,Morley96b}.  

Regardless of the model one chooses to examine, there is a common need to synthesise the microphysics with the global, \emph{collective} dynamics in order to accurately describe the observational data.  In order to produce a realistic model of the global dynamics, some subtle aspects of the microphysics need to be sacrificed.  Cellular automaton models of complex systems aim to employ only the dominant aspects of the microphysics to generate replicable and efficient automaton rules.  Such an approach was attempted by \cite{Morley96a} using a cellular automaton platelet model of mass accretion onto the stellar crust.  Although the microphysical details differ from those treated in this paper, the underlying philosophy of using a simple model to describe large scale behaviour is identical to ours.  We are similarly encouraged by the marked parallels between superfluid vortices and those that populate hard, type II superconductors.  The agreement between cellular automaton models of flux creep and vortex avalanches in superconductors and experimental data is excellent \citep{Linder,Jensen,Field,Bassler,Nicodemi01a,Nicodemi01b}, in particular with respect to the measurement of a power law over several decades in the avalanche (cf. glitch) size.

A general, first principles theory of self-organized criticality does not yet exist for any SOCS, let alone for pulsar glitches.  In this paper we look for empirical agreement between the output of our cellular automaton and the gross qualitative features in pulsar glitch data.  A detailed quantitative comparison between a first principles theory and data is not possible at this stage.

In this paper we present a cellular automaton model of pulsar glitches based on the mass unpinning model.  After describing the key features of a SOCS we revisit the observational pulsar glitch data in \S\ref{sec:SOCS}.  \S\ref{sec:unpinning} reviews the vortex unpinning paradigm and recent advances in the understanding of neutron star structure which are relevant to the paradigm.  \S\ref{sec:model} describes in detail the cellular automaton model and justifies physically the automaton rules.  In \S\ref{sec:results} the statistical behaviour of the model is explored with respect to driver parameters, internal physical parameters, and variations in the automaton rules.  Further discussion of the model's validity and a summary of our findings are contained in \S\ref{sec:conc}.

\section{Self-organized critical systems}
\label{sec:SOCS}
\subsection{General properties}
\label{subsec:gen_prop}

A system is described as a SOCS based on two underlying behavioural patterns.  \emph{Self-organized} implies the ability to develop structures and patterns in the absence of manipulation by an external agent.  \emph{Critical} implies that, like in phase transitions near the critical temperature, a small local perturbation can propagate throughout the entire system \citep{Jensen98}.

For a system to be in a self-organized critical (SOC) state it must first satisfy the following three conditions \citep{Jensen98,Melatos07a}:

\begin{enumerate}

\item\label{point:1}	It is composed of many discrete, mutually interacting elements, whose motions are dominated by local (e.g. nearest-neighbour) rather than global (e.g. mean-field) forces.

\item\label{point:2}	Each element moves when the local force exceeds a threshold.  In this way stress accumulates sustainedly at certain random locations (metastable stress reservoirs) while relaxing quickly elsewhere.

\item\label{point:3}	An external force drives the system slowly, in the sense that elements adjust to local forces rapidly compared to the driver time-scale.  Combined with local thresholds, this ensures that the system evolves quasistatically through a history-dependent sequence of metastable states.
\newline
\newline
\noindent If the above conditions are met, the following properties are generically observed \citep{Jensen98,Melatos07a}:
\newline
\item\label{point:4}	Transitions from one metastable state to the next occur via avalanches:  spatially connected chains of local equilibration events, in which one element relaxes and redistributes some local stress to its neighbors, which in turn can exceed their thresholds and relax (knock-on effect).

\item\label{point:5}	Avalanches have no preferred scale:  they can involve a few (commonly) or all (rarely) of the elements in the system.  Their sizes and lifetimes follow power law distributions, whose exponents are related.  Numerical values of the exponents depend on the spatial dimensionality of the system, the symmetry and strength of the local forces \citep{Field}, and the level of conservation \citep{Vespignani98,Pruessner02}.

\item\label{point:6}	Over the long term, the system tends to a critical state, which is stationary on average but fluctuates instantaneously.

\end{enumerate}

A driving force is not essential to the operation of a SOCS.  \cite{Pruessner02} assert that an external driving force compensates for a lack of conservation in the microscopic dynamics of a SOCS and hence is superfluous if the system is conservative on microscopic scales.  We claim that our model is locally non-conservative and hence a driving force is essential to maintaining criticality (see \S\ref{subsec:motion}).

Avalanches result from the relaxation of isolated metastable stress reservoirs.  The storage capacity of these reservoirs is scale invariant.  Each relaxation event is statistically independent from other relaxation events in size and when it occurs.  Two or more simultaneous relaxation events at different locations cannot be distinguished by their global effect (e.g. spin up of a neutron star), which is the cumulative result of multiple relaxations.  Likewise, relaxation of two or more smaller reservoirs causes an avalanche which is indistinguishable from that produced by the relaxation of a single large reservoir.  Statistical independence implies that the waiting times between avalanches follow Poisson statistics, and this is confirmed by models based on cellular automata.  

\cite{Alpar95,Alpar96} discussed the formation of depleted and capacitive regions of vortices in a neutron star interior, generated by steep gradients in the vortex pinning potential. Wherever the pinning potential is greater, the vortex density is also greater, resulting in a stronger local inter-vortex interaction (Magnus force).  When the Magnus force exceeds the pinning threshold, vortices are redistributed locally into neighbouring regions, relaxing the capacitive region (stress reservoir).  This qualitative picture bears the hallmark of a SOCS.

\subsection{Pulsar glitch statistics}
\label{subsec:puls_stats}

\begin{figure*}
\includegraphics[width = 10cm,angle=90]{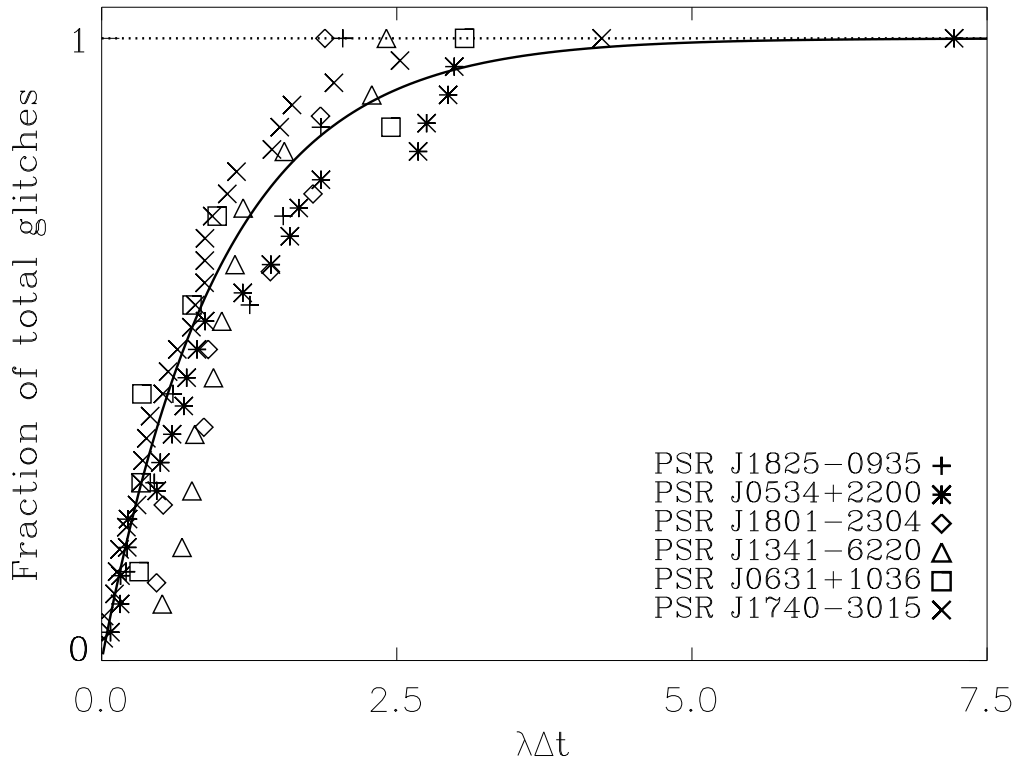}
\caption{Cumulative distributions of glitch waiting times $\Delta t$, for six pulsars with $N_{g} \geq 6$.  The horizontal scale $\lambda \Delta t$ is different for each pulsar; $\lambda$ is the mean glitch rate that minimises the K-S statistic for that pulsar as listed in Table \ref{tab:param}.  The solid curve represents the ideal, Poissonian, cumulative distribution, $P(\Delta t) = 1-e^{-\lambda\Delta t}$.  The waiting time distribution in an ideal SOCS follows this Poissonian.}
\label{fig:glitch_data_time}
\end{figure*}

\begin{table*}
\centering
\caption{Parameters of the glitch size and waiting time distributions for pulsars with $N_{\mathrm{g}}\geq 6$ \citep{Melatos07a}.  $a$ and $\lambda$ are chosen to minimise the K-S statistic.}
\begin{tabular}{l l l l l l l l}
\hline\hline
PSR J & $\lambda$ & $\lambda_{\mathrm{min}}$ & $\lambda_{\mathrm{max}}$  & $a$ & $a_{\mathrm{min}}$ & $a_{\mathrm{max}}$ & $10^{9}(\delta\nu)_{\mathrm{min}}$ \\
\hline
1825-0935 & 0.91 & 0.48 & 1.8 & 0.36 & $-$0.3 & 1.0 & 0.2 \\
0534+2200 & 0.91 & 0.57 & 1.3 & 1.2 & 1.1 & 1.4 & 1.1 \\
1801-2304 & 0.55 & 0.35 & 0.88 & 0.57 & 0.092 & 1.1 & 4 \\
1341-6220 & 1.8 & 1.2 & 5.6 & 1.4 & 1.2 & 2.1 & 10 \\
0631+1036 & 0.95 & 0.55 & 1.9 & 1.8 & 1.2 & 2.7 & 1.33 \\
1740-3015 & 1.5 & 1.2 & 2.5 & 1.1 & 0.98 & 1.3 & 0.7 \\
\hline
\end{tabular}
\label{tab:param}
\end{table*} 

\cite{Melatos07a} compiled all of the available glitch data for pulsars, analysing in detail the nine pulsars that have been observed to glitch at least five times.  We consider here the subset that have glitched more than six times ($N_{\mathrm{g}}\geq 6$).  Of the eight pulsars satisfying this criterion, six are consistent with a power law probability density function in the fractional glitch size $\delta\nu/\nu$, where $\nu$ is the spin frequency of the pulsar, implying a cumulative probability of the form
\begin{center}
\begin{equation}
\label{eq:power_law}
P(\delta\nu) =\frac{\left(\delta\nu\right)^{1-a}-\left(\delta\nu\right)_{\mathrm{min}}^{1-a}}
		{\left(\delta\nu\right)_{\mathrm{max}}^{1-a}-\left(\delta\nu\right)_{\mathrm{min}}^{1-a}} ,
\end{equation}
\end{center}
where $(\delta\nu/\nu)_{\mathrm{max}}$ is the maximum observed glitch size, and $(\delta\nu/\nu)_{\mathrm{min}}$ is the mimimum observed glitch size.  \cite{Melatos07a} found that $a$ takes a different value for each pulsar; when the data from all nine pulsars with $N_{\rm{g}}\geq 6$ are considered, Eq. (\ref{eq:power_law}) with a single value of $a$ is not a good description of the aggregate $\delta\nu/\nu$ distribution.  Two pulsars also exhibit quasiperiodic behaviour, which is also a natural manifestation of avalanche dynamics; when the global forces (driven externally) overwhelm the nearest-neighbour interactions, quasiperiodicity in the waiting time distribution is expected \citep{Jensen98}.  

For each of the six pulsars analysed here, we find all values of the exponent $a$, for which the data are not inconsistent with the analytic distribution at the $1 \sigma$ confidence level according to the Kolmogorov-Smirnov (K-S) test.  The range of $a$ is presented in Table \ref{tab:param}.  Table \ref{tab:param} also gives the exponent that minimises the K-S statistic, along with the smallest glitch $(\delta\nu/\nu)_{\mathrm{min}}$, for each pulsar \citep{Melatos07a}.  However, it should be noted that all values of $a$ within the $1 \sigma$ range (bounded by $a_{\mathrm{min}}$ and $a_{\mathrm{max}}$) are equally likely;  the exponent that minimises the K-S statistic is \textit{not} a best fit.  To appreciate why, suppose that the true underlying distribution of $\delta\nu/\nu$ is known for some pulsar.  Each time the pulsar glitches it samples this underlying $\delta\nu/\nu$ distribution.  The glitch data for this pulsar represent one possible realisation of the distribution containing $N_{\mathrm{g}}$ glitches.  There is no reason to attach greater significance to the $a$ value that minimises the K-S statistic with respect to the particular realisation than to any other in the $1\sigma$ range.

\cite{Melatos07a} also investigated the distribution of waiting times $\Delta t$, between successive glitches.  Based on the statistical independence of each avalanche and the assumption that the system is driven at a constant rate, it is expected that the probability density function for $\Delta t$ is Poissonian \citep{Jensen98}, implying a cumulative probability distribution of the form
\begin{center}
\begin{equation}
\label{eq:sum}
P(\Delta t) = \frac{1}{N_{\mathrm{g}}}\sum_{\Delta t_{\mathrm{min}}}
		\frac{\exp(-\lambda\Delta t_{\mathrm{min}})-\exp(-\lambda\Delta t)}{\exp(-\lambda\Delta t_{\mathrm{min}})-\exp(-\lambda\Delta t_{\mathrm{max}})} .
\end{equation}
\end{center}
The sum in (\ref{eq:sum}) is taken over the minimum observable waiting time $\Delta t_{\mathrm{min}}$, which is set by the gap between data spans in which a glitch is localised.  $\Delta t_{\mathrm{min}}$ is a function of the observing schedule and hence of epoch; effectively, therefore, it takes a different value for each glitch \citep{Melatos07a}.  $\Delta t_{\mathrm{max}}$ is the total interval over which a pulsar is observed.

Figure \ref{fig:glitch_data_time} shows cumulative histograms of waiting times $\Delta t$, for the same six pulsars listed in Table \ref{tab:param}, plotted agaist $\lambda \Delta t$, where $\lambda$ is the Poisson rate for each pulsar in units of $\rm{yr}^{-1}$.  Once again, $\lambda$ is chosen to minimise the K-S statistic for definiteness, but any value in the range $\lambda_{\mathrm{min}}\leq\lambda\leq\lambda_{\mathrm{max}}$ would do just as well at the $1 \sigma$ confidence level.  The observational data is overlaid with the ideal cumulative Poissonian $1-e^{-\lambda\Delta t}$ for comparison, taking $\Delta t_{\mathrm{min}}=0$ and $\Delta t_{\mathrm{max}}\rightarrow\infty$; that is, we do not correct for data gaps and a finite total observation for simplicity, unlike in \cite{Melatos07a}.  The agreement with the universal Poissonian is good for these six objects, with $0.35\,\mathrm{yr}^{-1}\leq\lambda\leq 5.6\,\mathrm{yr}^{-1}$.  Observational confirmation that pulsar glitch waiting times obey Poisson statistics upholds the prediction that the glitches arise from the relaxation of metastable stress reservoirs isolated by relaxed zones.
%

The observed distribution of glitch durations cannot be discussed at this time, as very few glitches have been resolved temporally \citep{McCulloch90}.

\section{Vortex unpinning in a pulsar}
\label{sec:unpinning}
The mass unpinning paradigm of pulsar glitches assumes that the inner crust is a dense nuclear lattice permeated by a superfluid that is threaded by many quantised vortices.  The vortices interact repulsively via the Magnus force\footnote{The Magnus force is in fact a fictitious force, similar in nature to the Coriolis or centrifugal forces.  A free vortex, in the absence of external forces, moves with the local superfluid flow.  In order for a vortex to move with respect to the local flow, a force must be applied such that the total force per unit length of vortex line, $\rho\bf{\kappa}\times(\bf{v}_{\rm{L}}-\bf{v}_{\rm{s}})$, is sufficient to sustain the motion of the line $\bf{v}_{\rm{L}}$ with respect to the local flow.} (per unit length of the vortex line) given by
\begin{center}
\begin{equation}
\label{eq:mag}
\mathbf{F}_{\mathrm{M}}=\rho_{\mathrm{s}}\mathbf{\kappa}\times(\mathbf{v}_{\mathrm{L}}-\mathbf{v}_{\mathrm{s}})\ ,
\end{equation}
\end{center}
where $\rho_{\mathrm{s}}$ is the superfluid density, $\mathbf{\kappa}$ is a vector pointing along the vortex, whose modulus $|\mathbf{\kappa}| = h/2m_{\mathrm{n}}$ is the circulation around a single vortex ($m_{\mathrm{n}}$ is the neutron mass), $\mathbf{v}_{\mathrm{L}}$ is the vortex line velocity, and $\mathbf{v}_{\mathrm{s}}$ is the local bulk superfluid velocity.  If allowed to move freely, vortices organize into a rectilinear array spaced evenly according to Feynman's rule (vortex area density $n = 4\pi\nu/\kappa$).  In keeping with condition \ref{point:1} in \S\ref{subsec:gen_prop}, each vortex is considered a discrete element of a SOCS.  For the purpose of practical modelling, given that we are dealing with $\sim 10^{16}$ vortices in a neutron star, we group vortices into bundles (in internal equilibrium) and regard the bundles as discrete elements.  The velocity field generated by each vortex is inversely proportional to the distance from the vortex line, so local interactions dominate global forces.  

Condition \ref{point:2} in \S\ref{subsec:gen_prop} exactly describes the pinning of vortices to the pulsar crust, whether the pinning centres are nuclear lattice sites, defects [monovacancies or impurities \citep{DeB98}], or grain boundaries.  Condition \ref{point:2} does not rule out pinning of neutron and proton vortices in the neutron star's core. Vortices are unpinned by the bulk superfluid flow when the Magnus force exceeds the pinning force.  \cite{Jones97} suggested that vortex pinning is too weak to occur in the pulsar crust, and that strong magnetic interactions between neutron and proton vortices are a feasible alternative.  In the event that strong pinning is provided by monovacancies alone, \cite{Jones98a,Jones98b} calculated the monovacancy concentration needed to provide a pinning force greater than the Magnus force, and claimed that, for $\rho_{\mathrm{s}}\geq3.4\times10^{16}\,\mathrm{kg\,m}^{-3}$, the concentration is unphysically large.

The external driver fulfilling condition \ref{point:3} in \S\ref{subsec:gen_prop} is the accumulating Magnus force generated by the build up of rotational lag between the vortex lines and the superfluid flow due to the electromagnetic spin down of the star.  The effect of this slow build up is two-fold.  Firstly, the vortices tend to slowly move radially outward, as the system strives to maintain the Feynman vortex density \citep{Alpar84,Jones98a}.  Secondly, abrupt adjustments occur when locally the pinning force threshold is exceeded.  Once a vortex is unpinned, it readjusts rapidly, either by returning to the `soup' of unpinned vortices (remembering that only $\sim 1\%$ of vortices are pinned at any one time), or by repinning at a nearby lattice or defect site.  Most mesoscopic vortex models assume that the inertia of vortex lines is negligible: once unpinned, vortices move immediately with the bulk superfluid flow, regardless of the direction of the unpinning force \citep{Donnelly}.  \cite{Field} pointed out that, at least in the context of flux creep in superconductors, such an assumption leads to `stick-slip' dynamics that are more likely to resemble those of an ideal sandpile (the quintessential SOCS), because inertial effects do not dominate the effect of the threshold.  The time-scale of the readjustment is much shorter than the driving time-scale of the stellar spin down.


Throughout this paper we assume that superfluid vortices are straight and parallel to the axis of rotation.  \cite{Jones98b} argued that the high, but finite, tension in vortex lines justifies this assumption, as the displacements caused by the pinning centres are much smaller than the nuclear separation.  Contrarily, it is the finite tension in vortices that allows them to be drawn into the potential well of the pinning centres and pin at all \citep{Jones98b}. \cite{Link02} suggested pinning centres in the nuclear lattice effectively span approximately thirty lattice sites, due to finite tension, where the internuclear spacing is $\sim 10^{-14}\rm{\,m}$.  Relative to the coarse graining scale of a practical cellular automaton (linear cell dimension up to $\approx 2\times10^{3}\rm{\,m}$), this deviation from straight vortices is clearly insignificant.  However, global hydrodynamic simulations suggest that the array breaks up into a vortex tangle above a critical rotational lag, driven by the Donnelly-Glaberson instability, with a net polarisation parallel to the rotation axis [see for example \cite{Peralta05,Peralta06b,Peralta06a}].

\section{Cellular automaton}
\label{sec:model}

\begin{figure*}
\includegraphics[width = 14cm]{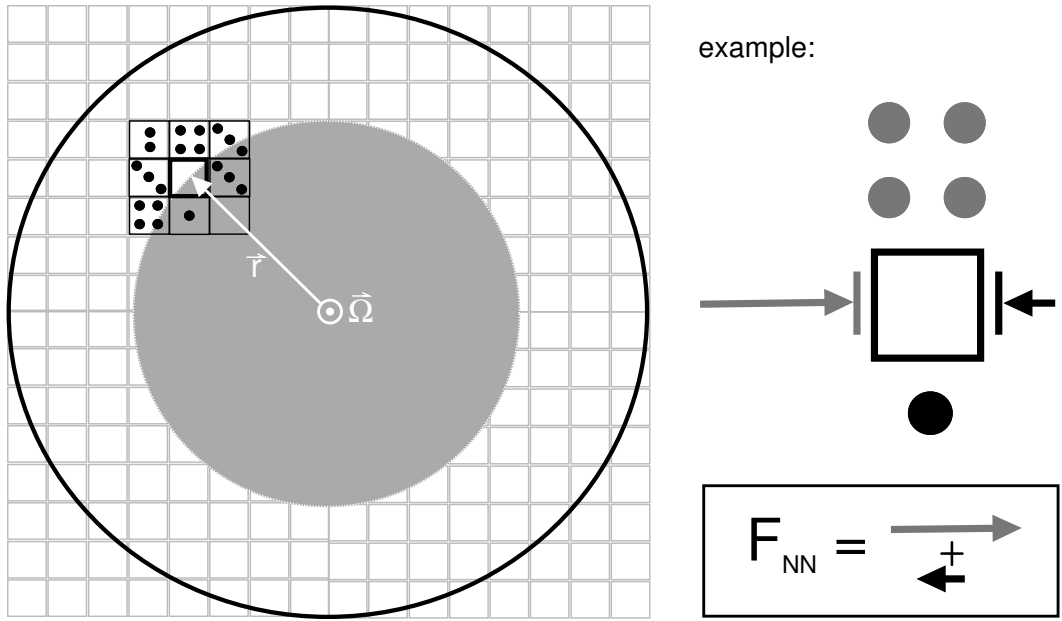}
\caption{Schematic of the cellular automaton.  The left side of the figure is a cartoon of the grid, illustrating how the force imbalance on a cell is calculated.  At the cell centred on the tip of the white arrow, the superfluid velocity field is calculated as the sum of three components:  the circular velocity due to the the pinned vortices residing in the grey shaded circle; the circular velocity due to the unpinned vortices in the grey region; and the summed velocity field due to vortex bundles in the eight nearest-neighbour and next-nearest-neighbour cells (bundles are represented as black dots).  The right side of the figure describes how the force imbalance due to the nearest-neighbour vortices, $F_{\rm{NN}}$, is calculated.  Each vortex bundle generates a velocity field according to a \emph{right-hand rule}, with the circulation vector pointing out of the page, which falls off inversely with distance from the centre of the bundle.}
\label{fig:grid}
\end{figure*}

The cellular automaton model presented in this paper takes the basic physical features of the inner crust-superfluid system and interprets them as simple rules.  Much of the microphysics that has emerged from detailed studies of pinning strengths, crustal structure, and turbulent flow (see references in \S\ref{sec:intro}) has not been taken into account directly when constructing the automaton rules, but it is relevant for understanding how the system can be `scaled up' (renormalised) to the necessary coarse grain.  In terms of reproducing the large-scale, collective dynamics of the SOCS, these simple cellular automaton rules have many advantages.  In particular, the simple rules allow for a large range of system sizes and parameter space to be modelled for long enough to obtain stationary statistics, without prohibitive computational cost.  Table \ref{tab:def} summarizes the nomenclature used in this and subsequent sections.

\subsection{Grid}
\label{subsec:grid}
The simulation grid contains $N_{\mathrm{grid}}^{2}$ cells, covering a circular `star' of radius $R$, representing a projection onto the equatorial plane.  The grid points are arranged in a rectilinear (square) array in order to easily identify nearest neighbours and to ensure that all cells are equal in size.  In reality, the equilibrium configuration of superfluid vortices is a triangular Abrikosov lattice.  For the purposes of this model, we assume that the two configurations are indistinguishable when vortices are bundled in large groups.  Vortex positions are restricted to the coordinates of the centre of each grid cell. As mentioned in \S\ref{sec:unpinning} there is an inherent assumption that the vortices are straight and parallel to the axis of rotation, and that the vortex dynamics are independent of the vortex length.  Indeed, the intervortex force is calculated as a force per unit length.

The number of vortices that occupy the grid is calculated using the Feynman relation
\begin{center}\begin{equation}
\label{eq:Feyn}
\oint \mathbf{v}_{\mathrm{s}}\cdot \mathbf{dl} = \kappa N_{\mathrm{v}}\ ,
\end{equation}\end{center}
where the integral is taken around a closed path (e.g. a circle of radius equal to the stellar radius), and $N_{\rm{v}}$ is the total number of vortices enclosed by the path.  In the limit of an infinite vortex distribution, the equilibrium distribution is a rectilinear array.  Although we have a finite array of vortices, we assume that they form a rectilinear array, such that the total number of vortices is
\begin{center}\begin{equation}
N_{\mathrm{v}}(R) = 4\pi^{2}\nu R^{2}/\kappa\ .
\label{eq:totv}
\end{equation}\end{center}

Once we know how many vortices should occupy the system as a whole, we bundle them such that, on average, there is one bundle per cell. Partly, this is done to reduce the number of discrete elements and keep the computation practical.  However, it is also done because uncertainties in pulsar timing limit the smallest glitch we can observe (corresponding to one bundle).  For any given pulsar we model the observed glitch range:  the minimum glitch size $G_{\mathrm{min}}$ (what constitutes a glitch is discussed in \S\ref{subsec:glitch}) occurs when one cell unpins, and the maximum $G_{\mathrm{max}}$ occurs when all cells unpin.  This restriction ensures that our model does not produce glitches outside the observationally imposed range.  This is in contrast with the model of \cite{Morley96a} in which the minimum glitch size is set by timing noise and the maximum by the total available energy.  Alternatively, we can have an average of $B\geq1$ bundles per cell.  In this scenario, the minimum glitch size is less than the minimum observed glitch size, and hence we can disregard all glitches  resulting from the unpinning of fewer than $B$ vortex bundles.  We choose our grid size such that 
\begin{center}\begin{equation}
\label{eq:grid_size}
N_{\mathrm{grid}}^{2} = \frac{G_{\mathrm{max}}}{G_{\mathrm{min}}}\ .
\end{equation}\end{center}
Pulsar timing data suggests, via (\ref{eq:grid_size}), a range of grid sizes varying from $N_{\mathrm{grid}}=10$ to a maximum of $N_{\mathrm{grid}}=100$\ \footnote{$N_{\mathrm{grid}}^{2}$ is the total number of cells in the square grid.  However, only cells that lie within the stellar radius are actually available.}.  The number of vortices per bundle is then 
\begin{center}\begin{equation}
N_{\mathrm{bun}}=N_{\mathrm{v}}/\left(BN_{\mathrm{grid}}^{2}\right) .
\end{equation}\end{center}

In an astrophysical context, however, we cannot assume that glitches smaller than the observable minimum do not take place, and with equal or greater frequency.  If the system is in a self-organized critical state, we expect glitches to occur at all scales, within the bounds of the system.  To resolve smaller $(\delta\nu/\nu)_{\mathrm{min}}$ in the model, we must fractionate into smaller bundles.  In order to make comparisons with the observational data, we preclude \emph{a priori} these smaller glitches from occurring in our model.  If the system is truly scale invariant, this effective window function (coarse-grain scale) should not interfere with its overall statistical behaviour.  In the flux creep model for superconductors developed by \cite{Jensen}, each cell holds at most one particle.  In our model each cell holds as many bundles as the avalanche history dictates. \cite{Bassler} take a similar approach, allowing many point pins in an extended cell.  

\subsection{Initial conditions}
\label{subsec:initial}
Initially the vortex bundles are laid out at random\footnote{Vortex bundles are placed at a randomly selected cell, one at a time, until all $N_{\mathrm{bun}}$ bundles have been distributed.} such that the mean occupancy of a cell is $B$ bundles.  Whether or not it is realistic to have a vortex distribution that is inhomogoneous on the scale of the grid cells ($\sim2\times10^{3}$m) depends upon the distribution of the pinning centres.  This point is discussed in more detail in \S\ref{sec:conc}.  From the glitch data analysed by \cite{Melatos07a}, we hypothesize that our system is in a SOC state at any given epoch, with an inhomogeneous configuration of pinned vortices, and take this to be the initial state of the automaton.  We do not model directly how this inhomogeneous state arises from an initially homogeneous state because the latter configuration is inaccessible to observations; it is disrupted by avalanches almost immediately after the neutron star is born.  Given that the random initial conditions of the model do not necessarily define a critical state of the system, we allow the simulation to complete many ($\sim 10^{4}$) time steps before considering the output, to ensure that criticality has been established.

\subsection{Pinning threshold}
\label{subsec:thresh}
The strength with which vortices are pinned to the stellar crust, $F_{\rm{thresh}}$, plays an essential role in the mass unpinning model of pulsar glitches.  Recent calculations imply that pinning to nuclear lattice sites is strongest in regions of low ($\sim 4.9\times 10^{15}\,\rm{kg\,m}^{-3}$) and high density ($\sim 1.6\times 10^{17}\,\rm{kg\,m}^{-3}$), with a maximum energy of $E_{\mathrm{p}}\sim 3$ MeV \citep{Avogadro}.  We take the pinning force to be the same everywhere on the grid, with value $E_{\mathrm{p}}/\xi$, where $\xi\sim 10^{-14}$m is the superfluid coherence length \citep{Elgaroy01}; i.e. $F_{\mathrm{thresh}}\sim 5\times10^{15}\mathrm{J\,m}^{-1}$.  For the purposes of the model we convert this to a threshold on the velocity lag, $\Delta \mathbf{v_{\mathrm{pin}}} = (\mathbf{v_{\mathrm{L}}}-\mathbf{v_{\mathrm{s}}})_{\rm{pin}}\sim10^{4}\mathrm{m\,s}^{-1}$.

Vortex pinning at lattice defects is also possible.  \cite{Link02} argued that if the pinned (coupled) fluid makes up $1\%$ of the total moment of inertia, then pinning must occur throughout the crust, suggesting that the crust should be treated as an amorphous solid with random pinning sites.  The randomness ensures that the contributions to the force of attraction towards the nuclear sites on either side of the vortex line do not cancel \citep{Jones98a}.  Another possibility is to adopt the random pinning potential used to model flux creep in superconductors \citep{Bassler}. 

Vortices unpin when the relative velocity of the superfluid and the vortex lines is large enough to produce a Magnus force $F_{\mathrm{M}}$ greater than the pinning threshold $F_{\mathrm{thresh}}$.  If the Magnus force does not exceed $F_{\mathrm{thresh}}$, the pinning force adjusts to balance the Magnus force exactly.  Given that the vortices are assumed to be inertialess, the excess force beyond the pinning threshold is of no relevance, as once the vortices unpin they flow with the ambient superfluid:
\begin{center}\begin{equation}
 F_{\mathrm{pin}}
 = 
 \left\{ \begin{array}{ll}
  F_{\mathrm{M}} & \textrm{if } F_{\mathrm{M}}<F_{\mathrm{thresh}}~\rm{,}\\
  0 	     & \textrm{otherwise}~.
\end{array} \right.
\end{equation}\end{center}

In our model, the superfluid velocity at a given point is the sum of three components:  the contribution from all pinned vortices that lie on or within the star-centred circle passing through the point (these vortices are assumed to be in a regular Feynman array on average); the contribution from all of the unpinned vortices lying within the above circle (also assumed to be in a regular Feynman array; i.e. mean-field approximation); and the contribution from the eight nearest-neighbour cells on the grid, as depicted in Fig. \ref{fig:grid}:
\begin{center}\begin{equation}
\mathbf{v}_{\mathrm{s}} = \mathbf{v}_{\mathrm{s,pinned}}+\mathbf{v}_{\mathrm{s,unpinned}}+\mathbf{v}_{\mathrm{s,NN}}\ .
\label{eq:deltav}
\end{equation}\end{center}
The first two contributions, hereafter named the mean-field terms, are calculated using Eq. (\ref{eq:Feyn}), assuming that the velocity field is exactly tangential to the closed path (a good approximation which becomes exact in the thermodynamic limit $N_{\mathrm{v}}\rightarrow\infty$).

By assuming that the vortices are arranged in a regular array, we can evaluate (\ref{eq:Feyn})  with tangential $\mathbf{v}_{\mathrm{s}}$, without knowing the location of each vortex (the validity of this approach is verified in \S\ref{subsec:gen}).  The Magnus force is then calculated from Eq. (\ref{eq:mag}), with $\mathbf{v}_{\mathrm{L}}$ equal to the veloicty of the stellar crust and $\mathbf{v}_{\mathrm{s,tot}}$ given by Eq. (\ref{eq:deltav}).  Initially, the mean Magnus force is very small.  As the star spins down, however, the pinned contribution does not diminish, while the unpinned contribution does, and the mean $\Delta\mathbf{v}$ increases.  

We emphasise that the automaton represents a locally averaged model where each cell covers many pinning sites.  Most of these sites are unoccupied because, for pinning at lattice defects (cf. seismic faults), the total number of vortices is much less than the number of pinning sites.

\subsection{Driver}
\label{subsec:drive}

In this  paper we consider two separate driving forces.  Over long time-scales ($>\delta\nu/\dot{\nu}$), the subject of \S\ref{subsec:long_term}, the Magnus force ramps up as the lag builds up between the pinned vortices and the unpinned superfluid.  Over short time-scales, we include thermally activated unpinning of the vortex bundles in random cells.\footnote{The contents of a randomly selected cell are deposited in the eight neighbouring cells.  A more realistic implementation may be to select the thermally unpinned cell only from those cells that are already close to the pinning threshold, and therefore most likely to unpin thermally.}  In both scenarios (short and long time-scales), small variations in $\Delta\mathbf{v} = \mathbf{v}_{L}-\mathbf{v}_{s}$ due to the eight nearest-neighbour cells trigger avalanches in the system.  By treating the nearest-neighbour interactions with greater precision than the longer range contributions, we are emphasizing the property outlined in condition \ref{point:1} of \S\ref{sec:SOCS}.

\subsection{Operational definition of glitches}
\label{subsec:glitch}
In the automaton model a glitch corresponds to the unpinning of vortex bundles.  Its size is determined by the total number of bundles (and hence vortices) unpinned.  The associated fractional spin up of the pulsar is then given by the fraction of the total moment of inertia carried by the recently unpinned fluid.  For example, if each vortex bundle represents the unpinning of fluid carrying $10^{-7}$ of the moment of inertia of the system, then an avalanche of ten bundles results in a glitch of size $\delta\nu/\nu = 10^{-6}$.

In practice, the angular momentum transferred to the crust by vortex unpinning depends on the product of the number of vortices unpinned (equivalently, the reduction in superfluid rotation rate) and the moment of inertia of the region through which they move before repinning, or until the avalanche stops, reaching a new quasi-equilibrium state.  The reason for this is found in the Onsager-Feynman relation (\ref{eq:Feyn}).  The superfluid velocity at some distance from the rotation axis is proportional to the number of vortices enclosed, so the further the unpinned vortices travel radially, the more the superfluid decelerates, and hence the more angular momentum is transferred to the crust.  The definition of glitch size given in this paper is therefore an approximation, valid only when three conditions are met: (i) all vortices travel the same radial distance during an iteration of the automaton; (ii) avalanches peter out after travelling one or two grid cells radially; and (iii) there are no steep radial gradients in vortex density (trivially satisfied by our uniform automaton, but not necessarily realistic).  We show in \S\ref{subsub:pin} that these conditions are fulfilled by our automaton primarily because the unpinning activity is restricted to a narrow annulus.  However, a more sophisticated automaton which incorporates radial gradients is needed to address this issue properly.

To facilitate future comparisons between the model and data, we clarify that the automaton describes spin-up events where the jump in spin frequency is unresolved by current timing experiments; the maximum rise time consistent with existing data is $\sim$40\,s \citep{Dodson02}.  Such glitches are accompanied by several recovery time scales, the longest of which may extend to the next glitch, together with a simultaneous jump in spin-down rate.  The automaton applies to this sort of unresolved spin-up event, although it does not seek to model the recovery physics.  Within the database of observed glitches, there are some glitches that do not conform to all facets of the above definition.  In general, these nonstandard glitches are observed in pulsars that have only glitched one or two times.  They are not described by the automaton in its current form, but they are usually large and hence relatively rare, so their impact on the statistics is muted.  Another kind of nonstandard glitch, which we make no attempt to model, is the time-resolved secondary spin up (`aftershock') seen in the Crab $20$ to $40$ days after a standard glitch \citep{Wong01}.

\subsection{Automaton rules}
\label{subsec:rules}
The rules of our cellular automaton encompass the key features of SOC outlined in \S\ref{sec:SOCS}:  discrete elements dominated by local forces, a threshold above which the elements move, and a slow driving force.  The algorithm that governs the dynamics of the model can be summarised as follows:
\begin{enumerate}
\item\label{step:1}The grid is initialised by randomly placing bundles of vortices on the cells that lie within the star, as in \S\ref{subsec:grid} and \S\ref{subsec:initial}.
\item\label{step:2}The force imbalance $\mathbf{F}_{\mathrm{M}}-\mathbf{F}_{\mathrm{thresh}}$ is calculated at each cell based on the scheme outlined in \S\ref{subsec:thresh}.  Cells where the force imbalance is positive are flagged as supercritical.
\item\label{step:3}Half the vortex bundles in a supercritical cell are repositioned on the adjacent cell that lies in the path of the bulk superfluid velocity vector, which may lie on the diagonal, leaving the supercritical cell with half its original contents.  Steps \ref{step:2} and \ref{step:3} are performed simulataneously, so that no bias arises from the order in which the cells are considered.
\item\label{step:4}The number of vortex bundles unpinned is recorded cumulatively.
\item\label{step:5}Steps \ref{step:1}--\ref{step:4} are repeated until no cells are supercritical. 
\item\label{step:6}The total number of vortex bundles unpinned is recorded as the glitch size for the current \emph{big} time step (where `big' is defined below).
\item\label{step:7}\begin{enumerate}
\item\label{step:7a}\emph{Either} the vortex bundles occupying a randomly selected cell are redistributed into the neighbouring cells; \emph{or}
\item\label{step:7b}the spin frequency of the crust (and hence $\mathbf{v}_{\mathrm{L}}/(2\pi R)$) is decremented by $\dot{\nu}\Delta t_{\rm{ts}}$, the number of unpinned vortices is proprotionally reduced, and the number of pinned vortices remains unchanged.
\end{enumerate}
\item\label{step:8}Steps \ref{step:1}--\ref{step:7} are repeated for a predetermined number of time steps, corresponding to either an arbitrary period of time [if \ref{step:7a} is used] or the number of time steps multiplied by $\Delta t_{\mathrm{ts}}$ (where $\Delta t_{\mathrm{ts}}$ is the length of each time step), if \ref{step:7b} is used.
\end{enumerate}
Throughout this paper we refer to the completion of steps \ref{step:1}--\ref{step:4} as a \emph{small} time step, and steps \ref{step:1}--\ref{step:5} as a \emph{big} time step.  Roughly speaking, a small time step lasts for the time it takes a vortex to be advected from one cell to its neighbour by the local superfluid flow.  Of course, this varies somewhat with position.  A big time step corresponds roughly to the mean waiting time between glitches, but please note that an avalanche does not always occur at every big time step given the above rules.  The total duration of the simulation depends on whether the driver is thermally activated vortex creep ($\sum\Delta t_{\rm{ts}}\ll\nu/\dot{\nu}$), or electromagnetic spin down ($\sum\Delta t_{\mathrm{ts}}\geq\nu/\dot{\nu}$) (see \S\ref{subsec:thresh}).

\subsection{Vortex motion}
\label{subsec:motion}

The rules for moving vortices to an adjacent cell following an unpinning event rely on the (incorrect) assumption that $|\mathbf{v}_{\mathrm{s}}|$ is the same in the vicinity of all supercritical cells, such that the time taken to travel from one pinning site to the next is the same throughout the grid.  That is, the discrete nature of the automaton means that the small time step sets the characteristic inter-cell travel time.  This approximation is justified in two ways in the model.  First, as long as the time to move one cell is much shorter than the driver time-scale, it is reasonable to set the small time step using the slowest vortices.  If, as we claim in \S\ref{subsec:gen}, most of the avalanche activity takes place in a thin annulus, this approach is accurate, because $|\mathbf{v}_{\mathrm{s}}|$ is roughly uniform over the annulus.  Second, only half the vortices in a supercritical cell are redistributed to its neighbours.  The travel time for vortices at the far side of the supercritical (departure) cell (relative to the destination cell) is greater than the travel time for vortices starting near the border.  Crudely, if vortices are evenly spaced throughout the departure cell, about half of them have sufficient time to reach the destination cell during a small time step.  As a corollary, the model contains a dissipative mechanism, which ensures that avalanches eventually terminate.  This feature means that our system is non-conservative.  Following \cite{Pruessner02}, we require an external driving force to maintain the system in a critical state.


\section{Statistics}
\label{sec:results}

\subsection{General features}
\label{subsec:gen}

\begin{table}
\centering
\caption{Default physical and computational parameters used to analyse the model.}
\begin{tabular}{l l l l l l l l}
\hline\hline
Parameter & Symbol & Default value  \\
\hline
Grid dimension & $N_{\rm{grid}}$ & 100 \\
Stellar radius & $R$ & $10\,\rm{km}$ \\
Initial spin frequency & $\nu_{0}$ & $30.25\rm{\,Hz}$ \\
Current spin frequency & $\nu$ & $13.90\rm{\,Hz}$ \\
Spin down rate & $\dot{\nu}$ & $0\rm{\,Hz\,s}^{-1}$ \\
Total vortices in simulation & $N_{\rm{v}}$ & $4\pi^2\nu R^2/\kappa$\\
Vortices per bundle & $N_{\rm{bun}}$ & $N_{\rm{v}}/N_{\rm{grid}}^2$ \\
Bundle factor (avg. bundles per cell) & $B$ & 1 \\
Pinning threshold & $F_{\rm{pin}}$ & $10^{4}\rm{\,m\,s}^{-1}$ \\
Pinned fraction & $\epsilon$ & 0.01 \\
\hline
\end{tabular}
\label{tab:def}
\end{table} 

Three quantities in particular are used to characterise our cellular automaton model of pulsar glitches:  the distribution of glitch sizes $\delta\nu/\nu$, durations $t$, and waiting times $\Delta t$.  The glitch duration is defined as the number of consecutive time steps at which there exist supercritical cells. 

The hypothesis that the automaton results in avalanche dynamics, driven slowly by electromagnetic spin down or random thermal unpinning, is now tested.  The model is allowed to run for $10^{5}$ big time steps.  In \S\ref{subsec:dyn}--\S\ref{subsec:nu} we consider time periods over which $\nu$ does not decrease significantly.  In \S\ref{subsec:long_term} we consider longer time periods over which spin down becomes important.  We characterise the model by exploring its response to changes in the controlling parameters, both physical and computational.  The physical parameters are the spin frequency $\nu$, the pinning threshold $\Delta \mathbf{v}_{\rm{pin}}$, and the fraction of vortices that are pinned $\epsilon$.  The compuational parameters are the numbers of bundles $N_{\rm{bun}}\equiv B N_{\rm{grid}}^{2}$ and grid cells $N_{\rm{grid}}^{2}$.

Fig. \ref{fig:compilation} shows the raw output of the model in the form of a time series of glitch sizes for a standard set of parameters (defined Table \ref{tab:def}).  Unless otherwise specified, these are the parameters used to generate the model output.  This particular set of parameters is chosen to balance the need for sufficient avalanches to generate good statistics versus computational efficiency.  The vertical axis gives the size in terms of the fraction of the total superfluid moment of inertia that unpins during the avalanche.  The inset zooms in on a segment of the time series spanning $10^{3}$ big time steps.  Although the simulation runs for a total of $10^{5}$ time steps (without spin down), the first $10^{3}$ time steps are discarded to allow the system to establish stationary statistics.  \cite{Jensen98} asserted that fractal dynamics necessarily arise in a SOCS.  Although we do not endeavour to quantify this mathematically, we are able to demonstrate, by considering time series such as the one shown in Fig. \ref{fig:compilation}, that our system produces self-similar dynamics; qualitatively, the time series looks the same whether we plot $10^{5}$ or $10^{3}$ big time steps.

The lower panel of Fig. \ref{fig:compilation} shows the probability density functions of glitch sizes $\delta\nu/\nu$, durations $t$, and waiting times $\Delta t$, plotted as histograms, for the standard set of parameters defined in the caption.  Both $\delta\nu/\nu$ and $t$ obey power law statistics, with probability density functions $p(\delta\nu/\nu)\propto (\delta\nu/\nu)^{-a}$ and $p(t)\propto t^{-b}$ respectively.  This is a characteristic feature of a SOCS (see \S\ref{subsec:gen_prop}).  For the standard parameters in Fig. \ref{fig:compilation}, we obtain $a=2.9$ and $b=4.8$.  More generally, for the parameter ranges studied in this paper, we find $2.0\leq a\leq 4.3$ and $2.2\leq b\leq 5.5$ (see \S\ref{subsec:dyn}--\S\ref{subsec:nu}).  In a SOCS, $a$ and $b$ are related by \citep{Jensen98} 
\begin{equation}
 b = 1+(a-1)\frac{\gamma_{2}}{\gamma_{3}}\rm{\ ,}
\end{equation}
such that the size of an avalanche scales with its linear extent $L$ and $(\delta\nu/\nu)\propto L^{\gamma_{2}}$ and the duration scales as $t\propto L^{\gamma_{3}}$.  The exponents in Fig. \ref{fig:compilation} imply $\gamma_{2}/\gamma_{3}\approx 2.0$, which is exactly what one expects for compact, two-dimensional avalanches with $\gamma_{2}=2$ and $\gamma_{3}=1$ (see \S\ref{subsec:long_term}).  Even more encouragingly, we find $(b-1)/(a-1)=\gamma_{2}/\gamma_{3}\approx 2.0$ in many of our numerical experiments where criticality applies, even though $a$ and $b$ individually cover wide ranges.

\begin{figure}
\includegraphics[width = 5.5cm,angle=90]{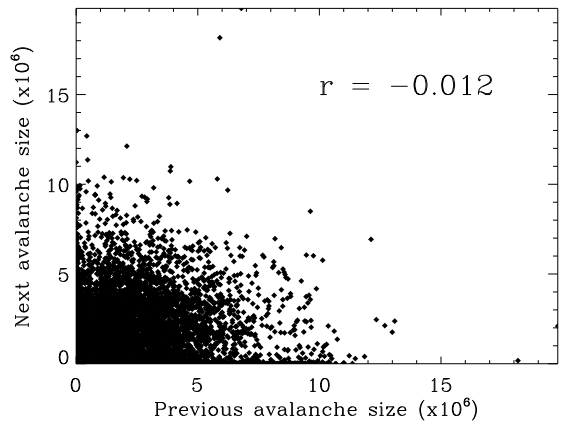}
\caption{Next avalanche size versus previous avalanche size for $9\times10^{4}$ avalanches.  $r$ is the linear Pearson correlation coefficient.  The automaton shows no sign of `memory' as $|r|\ll 1$.}
\label{fig:corr}
\end{figure}

Based on the data analyzed in \cite{Melatos07a}, as well as the likeness between our model and other cellular automata that display self-organized critical behaviour,  we expect the distribution of waiting times to be exponential, in keeping with a Poisson process (statistically independent avalanches).  The bottom right panel in Fig. \ref{fig:compilation} confirms this expectation:  the waiting-time probability density function is $p(\Delta t)\propto e^{-\lambda\Delta t}$, with $\lambda = 6.8\times 10^{-3}$ in units of an inverse big time step.  To further verify the property of statistical independence, we calculate the linear Pearson correlation coefficient relating the size of an avalanche and the size of the avalanche immediately preceding it, as plotted in Fig. \ref{fig:corr}.  We find that for $10^{5}$ avalanches, the correlation coefficient is $-0.012$, implying that there is almost no correlation between an avalanche and its predecessor. 

Contrary to the general idea that a SOCS tends naturally to its critical state, we are forced to fine tune the parameters to elicit avalanches.  The system is particularly sensitive to $\epsilon$ and $|\Delta \mathbf{v}_{\rm{pin}}|$.  Similarly, the exponents of the best power law fits to the glitch size and duration distributions (see \S\ref{subsec:dyn} and \S\ref{subsec:nu}) depend on the same tuned parameters.  This is also observed to be the case experimentally when superconducting vortices respond to a change in the ramp rate of the magnetic field \citep{Field}.  In cellular automaton models of forest fires, the number of trees between two ignitions needs to be tuned according to system size, to avoid a cutoff or bump in the size distribution \citep{Pruessner02}.

\begin{figure*}
\includegraphics[width = 6.2cm,angle=90]{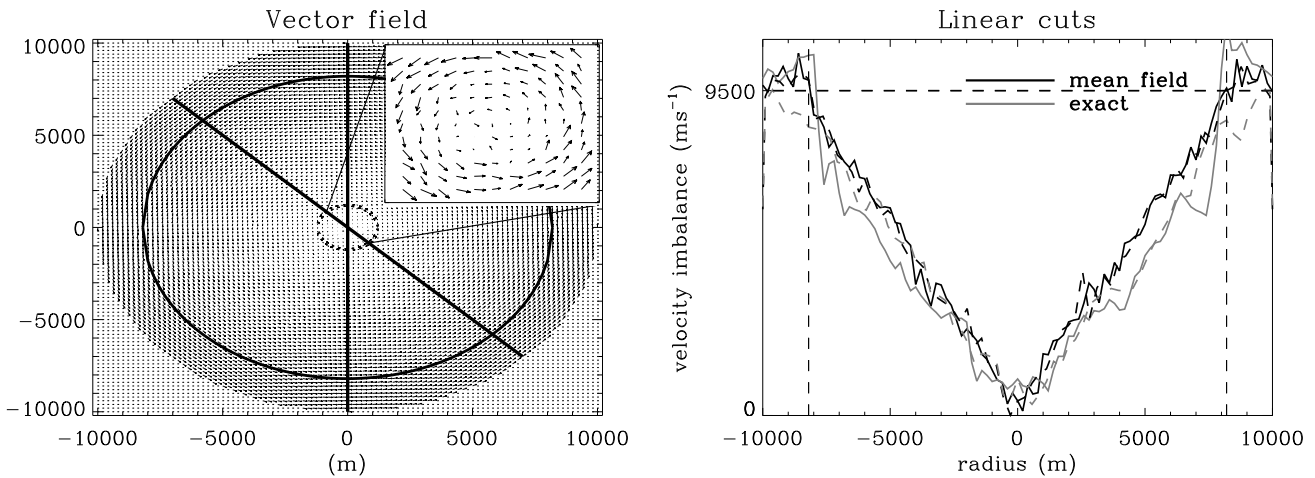}
\caption{\textit{Left:} A vector plot of the velocity imbalance $\Delta \mathbf{v}=\mathbf{v}_{\rm{L}}-\mathbf{v}_{\rm{s}}$, generated using the mean-field approximation described in Eq. (\ref{eq:deltav}).  The inset zooms in on the central $10\times10$ grid cells.  The solid black circle shows the position of the critical radius, where the mean-field contribution to $\Delta \mathbf{v}$ is approximately equal to $\Delta \mathbf{v}_{\rm{pin}}$. \textit{Right:}  Profiles of $\Delta \mathbf{v}$ along a straight-line cut through the grid, as a function of distance from the origin, as calculated by the mean-field (\textit{black}) and exact (\textit{N}-body) (\textit{grey}) algorithms.  The solid and dashed curves correspond to the vertical and diagonal cuts drawn in the left panel.  The two vertical dashed lines show the position of the critical circle.  The horizontal dashed line marks $\Delta\mathbf{v}_{\rm{pin}}$.}
\label{fig:velocity}
\end{figure*}

Before analysing the results produced by the cellular automaton, we assess the accuracy of the mean-field approximation outlined in \S\ref{subsec:thresh}.  Fig. \ref{fig:velocity} shows the velocity imbalance across the grid (as both a vector field and along a linear cut) generated using the mean-field approximation for a random distribution of vortex bundles (with a mean occupancy of one bundle per cell).  The result is compared with an exact, `\textit{N}-body' algorithm, in which $\Delta\mathbf{v}$ at any cell is computed by summing the $\mathbf{v}_{\rm{s}}$ produced at that cell by every bundle on the grid.  The exact algorithm is not used elsewhere in this paper due to computational cost.  Discrepancies between the mean-field and exact calculations are greater when a line is taken vertically through the velocity field, rather than diagonally.  This stems from the discrete, rectilinear nature of the grid.  The two calculations are also in greater agreement further from the origin.  By considering $\Delta\mathbf{v}$ in the vicinity of the \textit{critical circle}, where the mean-field contribution to the Magnus force approximately equals the pinning threshold (vertical dashed lines in right panel of Fig. \ref{fig:velocity}), we see that the nearest-neighbour and gridding fluctuations are comparable.  Yet the effect of gridding on the dynamics turns out to be negligible, because the output of the model is a power law in $\delta\nu/\nu$.  If the fluctuations were dominated by gridding effects, we would not expect power law statistics.  

The bold circle overlaid on the velocity field in Fig. \ref{fig:velocity} is called the critical circle. Well beyond this circle, the mean-field contribution consistently exceeds the pinning threshold, such that the nearest-neighbour variations are insufficient to alter the supercritical state of the cell.  Out here, vortex bundles never pin.  Similarly, well within this radius, the nearest-neighbour variations are insufficient to alter the subcritical state of the cell.  In here, vortex bundles never unpin.  Near the circle, however, the nearest neighbour fluctuations are pertinent; indeed, they cause the avalanches we observe.  The location of the `active' annulus thereby defined depends on the physical and the computational parameters that define the model.

\begin{figure*}
\begin{center}
\includegraphics[width = 8cm,angle=90]{fig_3a.epsi}
\end{center}
\includegraphics[width = 5.7cm,angle=90]{fig_3b.epsi}
\caption{\textit{Top}:  Automaton output for $9\times 10^{4}$ big time steps.  The radius of the star is $R=10$\,km, the unpinning threshold is $|\Delta\mathbf{v}_{\rm{pin}}| = 1.0\times10^{4}$\,ms$^{-1}$, the neutron superfluid density is $\rho_{\mathrm{s}}=10^{16}$\,kg\,m$^{-3}$, the initial spin frequency is $\nu_{0}=30.25$\,Hz, and the current spin frequency is $\nu = 13.9$\,Hz.  The avalanche size is given as the fraction of the total fluid moment of inertia that unpins during the avalanche.  The fraction of pinned vortices is taken to be $\epsilon = 0.01$.  The main panel displays a time series of avalanche sizes beginning after $10^{4}$ time steps elapse, in order to ensure that a self-organized critical state has time to emerge from the initially random conditions. The inset zooms in on $5\times 10^{3}$ time steps.  \textit{Bottom}:  Probability density funtions of $\delta\nu/\nu$ (\textit{left}), durations $t$ (\textit{centre}) and $\Delta t$ (\textit{right}) plotted as histograms on a log-log (\textit{left} and \textit{centre}) and log-linear (\textit{right}) scale, using $10^{2}$ logarithmically (\textit{left} and \textit{centre}) and linearly (\textit{right}) spaced bins.  The same parameters are used as for the top panel.  The size and duration distributions are fitted with power laws of the form $p(\delta\nu/\nu)=(\delta\nu/\nu)^{-a}$ and $p(\delta\nu/\nu)=(\delta\nu/\nu)^{-b}$ respectively.  The waiting time distribution is fitted with an exponential of the form $p(\Delta t)=e^{-\lambda\Delta t}$, where $\lambda$ is in units of the reciprocal of one big time step.  All fitted functions are plotted as dashed lines.}
\label{fig:compilation}
\end{figure*}

\begin{figure*}
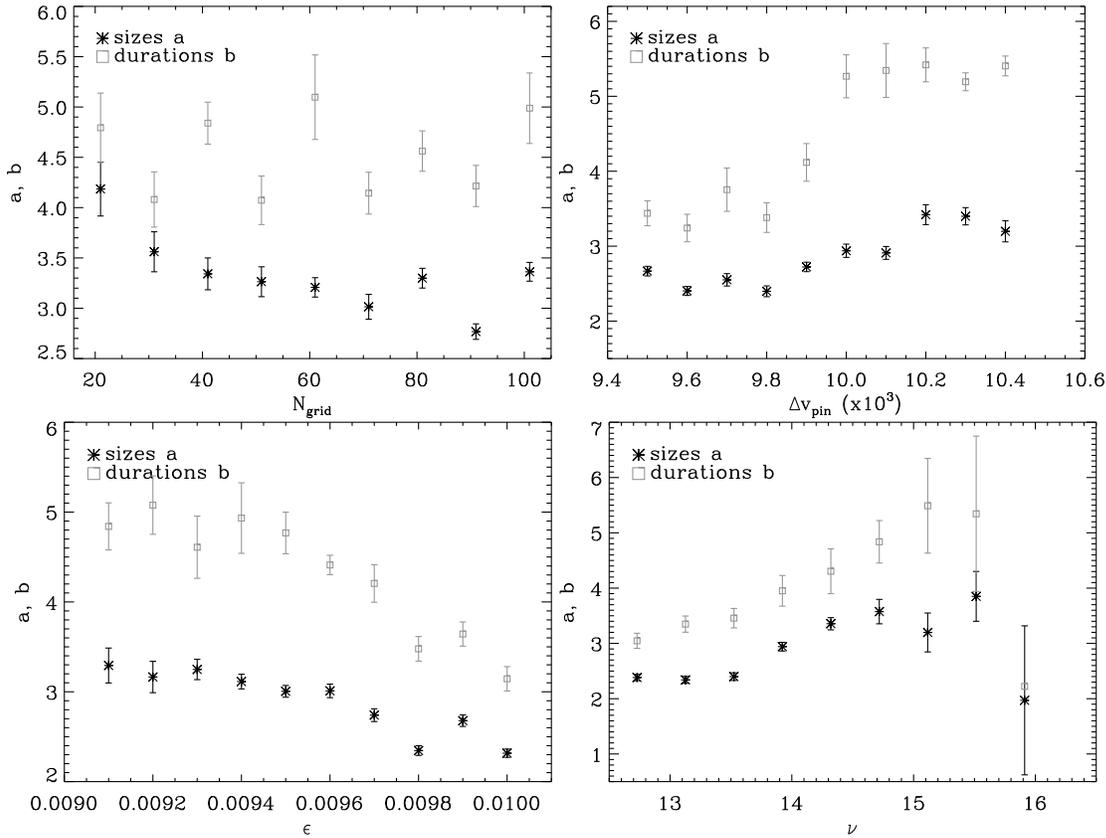

\includegraphics[width = 5.5cm,angle=90]{fig_5a.epsi}
\includegraphics[width = 5.5cm,angle=90]{fig_5b.epsi}
\includegraphics[width = 5.5cm,angle=90]{fig_5c.epsi}
\includegraphics[width = 5.5cm,angle=90]{fig_5d.epsi}
\caption{Fitted power law exponents for the distribution of glitch sizes ($\ast$) and durations ($\square$).  The error bars represent the $1\sigma$ uncertainty returned by the least-squares fitting algorithm.  \textit{Top left}:  $10\leq N_{\mathrm{grid}}\leq 100$.  \textit{Top right}:  $9.5\times 10^{4}\leq|\Delta\mathbf{v}_{\rm{pin}}|\leq 10.4\times 10^{4}\rm{\,m\,s}^{-1}$.  \textit{Bottom left}:  $0.0092\leq\epsilon\leq0.010$.  \textit{Bottom right}:  $12.73\leq\nu\leq15.91\,\rm{Hz}$.
Automaton parameters:  $N_{\mathrm{grid}}=100$, $B=1$, $10^{5}$ big time steps.
Physical parameters:  $\rho_{\rm{s}}=10^{16}\rm{\,kg\,m}^{-3}$, $\nu_{0}=30.25\rm{\,Hz}$, $R=20\rm{\,km}$.}
\label{fig:pl}
\end{figure*}

\subsection{Dynamic range}
\label{subsec:dyn}

\begin{figure*}
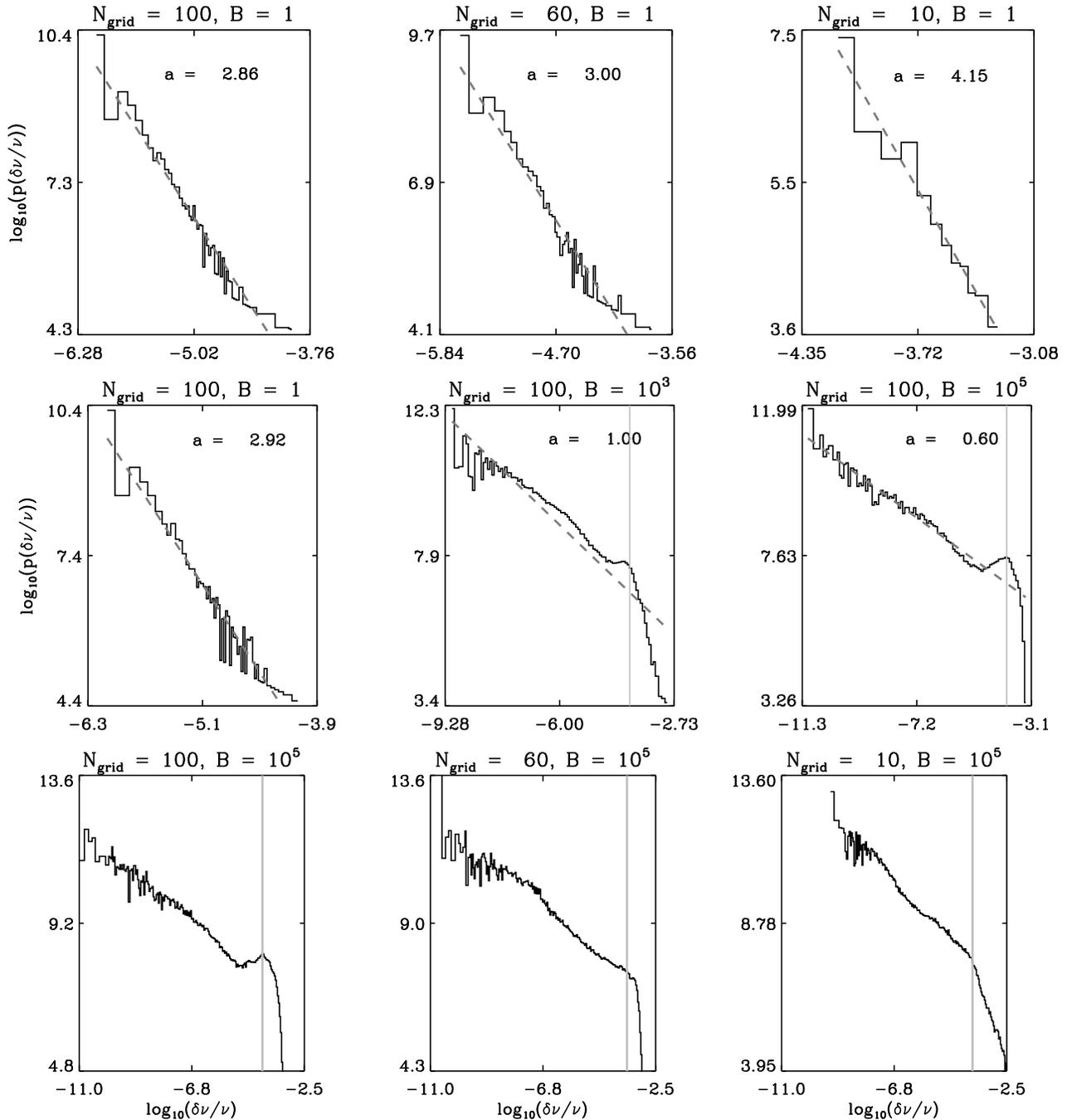

\includegraphics[width = 5.7cm,angle=90]{fig_7a.epsi}
\begin{center}
\includegraphics[width = 5.7cm,angle=90]{fig_7b.epsi}
\end{center}
\begin{center}
\includegraphics[width = 6.0cm,angle=90]{fig_7c.epsi}
\end{center}
\caption{Probability density functions $p(\delta\nu/\nu)$ of avalanche sizes $\delta\nu/\nu$, plotted as histograms on a log-log scale, using $10^{2}$ logarithmically spaced bins, for $10^{5}$ big time steps.  The histograms are fitted with a power law of the form $p(\delta\nu/\nu)\propto(\delta\nu/\nu)^{-a}$ (\textit{dashed lines}).  \textit{Top}:  Grids with $N_{\mathrm{grid}}=100$ (\textit{left}), $60$ (\textit{centre}) and $10$ (\textit{right}). All panels have bundle factor $B=1$. \textit{Middle}:  Bundle factors $B=1$ (\textit{left}), $10^{3}$ (\textit{centre}) and $10^{5}$ (\textit{right}). All panels have $N_{\rm{grid}}=100$.  \textit{Bottom}:  $B=10^{5}$ and $N_{\mathrm{grid}}=100$ (\textit{left}), $60$ (\textit{centre}) and $20$ (\textit{right}).  All panels have $B=10^{5}$.  The vertical grey line denotes the crossover scale.
Physical parameters:  $|\Delta\mathbf{v}_{\rm{pin}}|=10^{4}\rm{\,m\,s}^{-1}$, $\rho_{\rm{s}}=10^{16}\rm{\,kg\,m}^{-3}$, $\nu_{0}=30.25\rm{\,Hz}$, $\nu=13.9\rm{\,Hz}$, $\epsilon=0.01$, $R=20\rm{\,km}$.}
\label{fig:hist_N}
\end{figure*}

\begin{figure*}
\includegraphics[width = 5.7cm,angle=90]{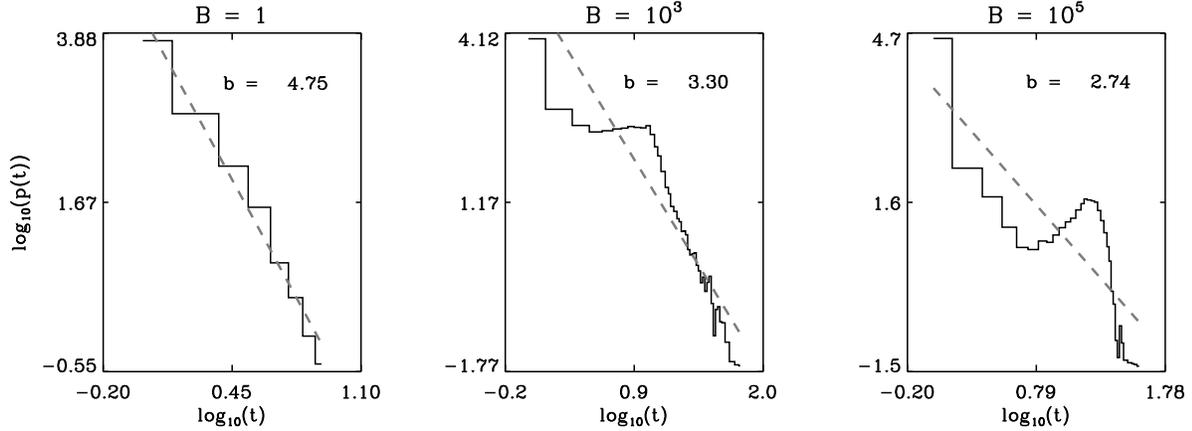}
\caption{Probability density functions $p(t)$ of avalanche durations $t$, for bundle factors $B=1$ (\textit{left}), $10^{3}$ (\textit{centre}) and $10^{5}$ (\textit{right}), plotted as histograms on a log-log scale, using $10^{2}$ logarithmically spaced bins.  The histograms are fitted with a power law of the form $p(t)\propto t^{-b}$ (\textit{dashed lines}). 
Automaton parameters:  $N_{\mathrm{grid}}=100$, $10^{5}$ big time steps.
Physical parameters:  $|\Delta\mathbf{v}_{\rm{pin}}|=10^{4}\rm{\,m\,s}^{-1}$, $\rho_{\rm{s}}=10^{16}\rm{\,kg\,m}^{-3}$, $\nu_{0}=30.25\rm{\,Hz}$, $\nu=13.9\rm{\,Hz}$, $\epsilon=0.01$, $R=20\rm{\,km}$.}
\label{fig:dur_B}
\end{figure*}

\begin{figure*}
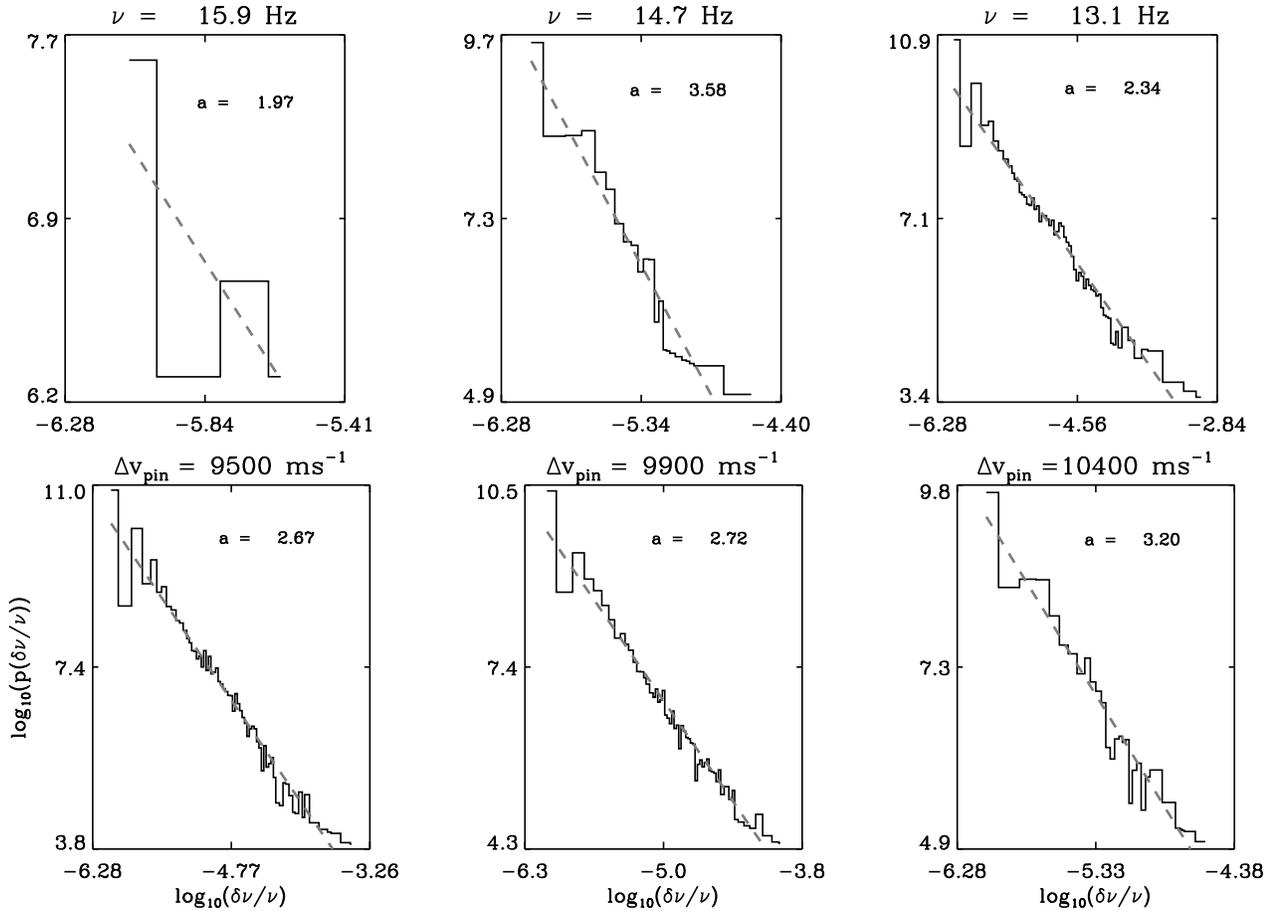

\includegraphics[width = 5.7cm,angle=90]{fig_9a.epsi}
\begin{center}
\includegraphics[width = 6.1cm,angle=90]{fig_9b.epsi}
\end{center}
\caption{Probability density functions $p(\delta\nu/\nu)$ of avalanche sizes $\delta\nu/\nu$, plotted as histograms on a log-log scale, using $10^{2}$ logarithmically spaced bins.  The histograms are fitted with a power law of the form $p(\delta\nu/\nu)\propto(\delta\nu/\nu)^{-a}$ (dashed lines).  \textit{Top}:  Current spin frequency $\nu=15.9$ (\textit{left}), $14.7$ (\textit{middle}) and $13.1\rm{\,Hz}$ (\textit{right}).  \textit{Bottom}:  Pinning threshold $|\Delta\mathbf{v}_{\mathrm{pin}}|=9.5\times 10^{3}$ (\textit{left}), $9.9\times 10^{3}$ (\textit{middle}) and $10.4\times 10^{3}\rm{\,m\,s}^{-1}$ (\textit{right}).
Automaton parameters:  $N_{\mathrm{grid}}=100$, $10^{5}$ big time steps, $B=100$.
Physical parameters:   $\rho_{\rm{s}}=10^{16}\rm{\,kg\,m}^{-3}$, $\nu_{0}=30.25\rm{\,Hz}$, $\epsilon=0.01$, $R=20\rm{\,km}$.}
\label{fig:hist_nu}
\end{figure*}

\begin{figure*}
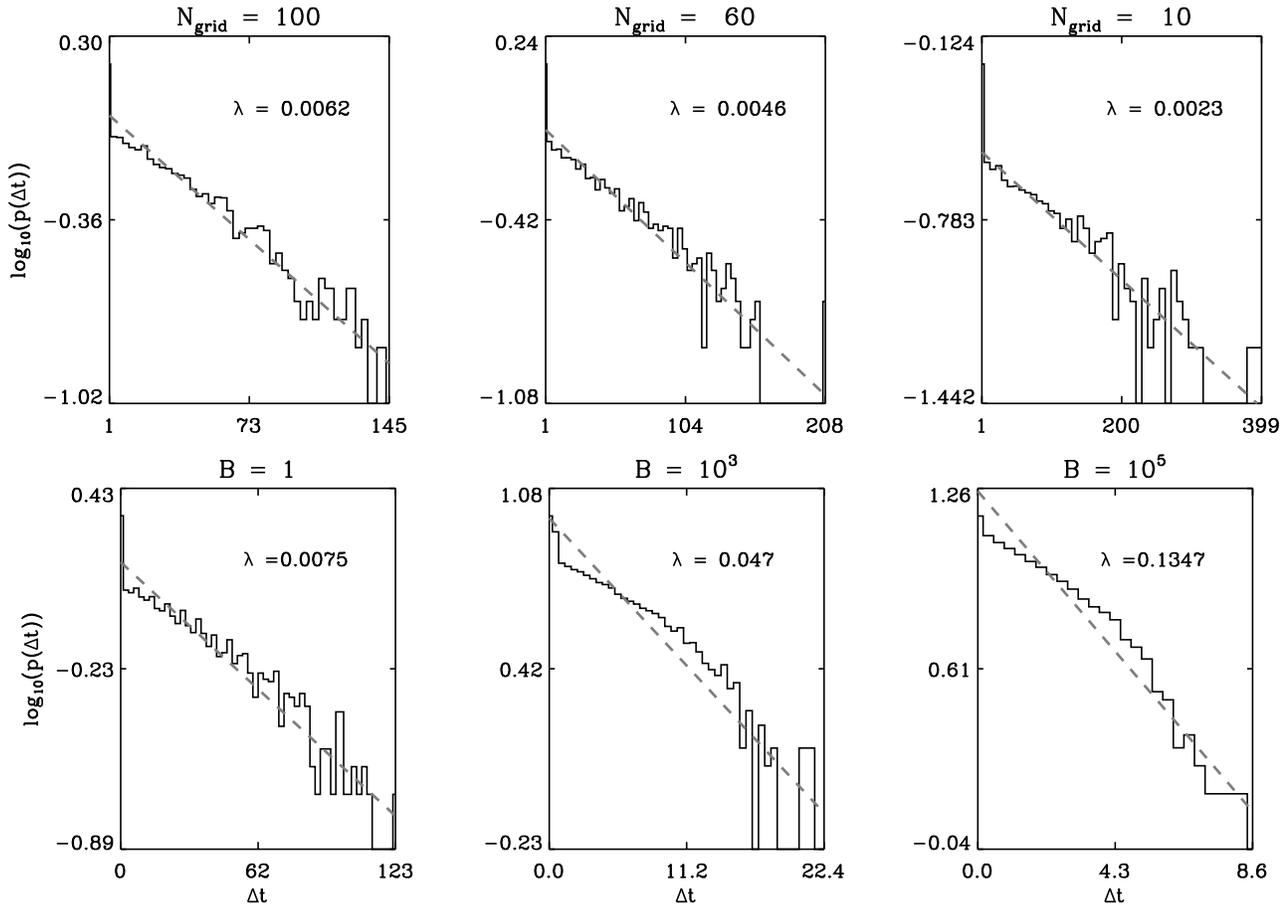

\includegraphics[width = 5.7cm,angle=90]{fig_10a.epsi}
\begin{center}
\includegraphics[width = 6cm,angle=90]{fig_10b.epsi}
\end{center}
\caption{Probability density functions $p(\Delta t)$ of waiting times $\Delta t$, plotted as histograms on a log-linear scale, using $10^{2}$ linearly spaced bins, for $10^{5}$ big time steps.  The histograms are fitted with an exponential of the form $p(\Delta t)\propto e^{-\lambda\Delta t}$ (dashed lines).  \textit{Top}:  Grid size $N_{\mathrm{grid}}=100$ (\textit{left}), $60$ (\textit{centre}) and $10$ (\textit{right}).  Bundle factor $B=1$.  \textit{Bottom}:  Bundle factor $B=1$ (\textit{left}), $5\times 10^{2}$ (\textit{centre}) and $10^{5}$ (\textit{right}).  Grid size $N_{\mathrm{grid}}=100$.
Physical parameters:  $|\Delta\mathbf{v}_{\rm{pin}}|=10^{4}\rm{\,m\,s}^{-1}$, $\rho_{\rm{s}}=10^{16}\rm{\,kg\,m}^{-3}$, $\nu_{0}=30.25\rm{\,Hz}$, $\nu=13.9\rm{\,Hz}$, $\epsilon=0.01$, $R=20\rm{\,km}$.}
\label{fig:wait_N}
\end{figure*}

In \S\ref{subsec:grid} we motivate the choice of grid size by referring to the observed dynamic range of pulsar glitches.  In any individual pulsar the maximum dynamic range of $\delta\nu/\nu$ is $10^{4}$ (for PSR J1704--3015), and the minimum is $10^{1}$ (for PSR J0537--6910 \footnote{This is one of two pulsars that glitch quasiperiodically; the other is Vela.}).  Hence, from (\ref{eq:grid_size}), we run our simulation on grids ranging from $N_{\rm{grid}}=10$ to $N_{\rm{grid}}=100$ (the physical size remains unchanged at $R=20\rm{\,}km$). By construction, the range of avalanches outputted by the model should not exceed $G_{\mathrm{max}}/G_{\mathrm{min}}$ (i.e. $10^{2}$ for the $10\times10$ grid, and $10^{4}$ for $N_{\rm{grid}}=100$ grid). Fig. \ref{fig:hist_N} confirms this expectation.  The top panels are histograms of the avalanches produced by systems with $N_{\rm{grid}}=100$, $N_{\rm{grid}}=60$ and $N_{\rm{grid}}=10$ over $10^{5}$ time steps.  The dynamic range is greatest for the largest grid ($10^{-6.1}\leq \delta\nu / \nu\leq 10^{-4.2}$) and smallest for the smallest grid ($10^{-4.2}\leq \delta\nu / \nu\leq 10^{-3.2}$). It is not expected that these dynamic ranges span the full range $G_{\rm{min}}\leq \delta\nu/\nu\leq G_{\rm{max}}$, as many of the grid cells lie inactive, well within the critical circle discussed in \S\ref{subsec:gen}.  It is, however, encouraging that the larger grid sizes do in fact produce a larger dynamic range.

Each of the histograms in Fig. \ref{fig:hist_N} is overlaid with a power law best fit of the form $p(\delta\nu/\nu)\propto (\delta\nu/\nu)^{-a}$, estimated using a least-squares algorithm, with the fitted exponent indicated on each plot.  By the central limit theorem, the uncertainty in each bin is inversely proportional to the square root of its contents.  The fitted exponent is graphed as a function of $N_{\rm{grid}}$ in Fig. \ref{fig:pl}, with error bars taken from the least-squares algorithm.  There is a trend in both $a$ and $b$ to increase with increasing $N_{\rm{grid}}$, in keeping with the increasing dynamic range with $N_{\rm{grid}}$.  When the simulation is run for $2\times 10^{5}$ time steps, the dynamic range does, as expected, increase; however, it still falls well short of $G_{\mathrm{max}}/G_{\mathrm{min}}$.  The number of bundles available to unpin is restricted to a small fraction ($\sim 15\%$) of $B N_{\rm{grid}}^{2}$ that lie in the vicinity of the critical circle.  To better match the desired dynamic range, we should select the grid size such that $G_{\mathrm{max}}/G_{\mathrm{min}}$ cells lie in the active zone near the critical circle.

The glitch sizes generated by the model should be treated as a indicative only.  Evidently, several of the pulsars analyzed by \cite{Melatos07a} exhibit glitches several orders of magnitude smaller than those produced by the automaton.  It is easy to elicit such small glitches from the automaton by adjusting $F_{\rm{pin}}$ or $\nu-\nu_{0}$ so that the active circle lies closer to the rotation axis, thereby enclosing fewer vortices.  We have not explored these possibilities here as there is too much freedom in the choice of parameters at present. (See \S\ref{subsub:frac} for further discussion.)  If pinning is strongest in two or more annuli within the crust, the model predicts that glitches fall into two or more size brackets, depending on where the annuli are located.  It should be noted, however, that \cite{Melatos07a} found no evidence for a bimodal distribution when analyzing the latest statistics from \emph{individual} pulsars, whose size distributions are consistent with unbroken power laws.

\subsection{Vortex bundling}
\label{subsec:bundle}

An underlying assumption of the model is that the scale on which the pinning centres and vortices are coarse grained does not alter the behaviour of the output qualitatively.  In this section, we explore the impact of maintaining the grid size whilst varying the mean number of bundles per cell $B$.  The simulation is designed such that the smallest (largest) avalanche occurs when one (every) bundle unpins.  In a scale invariant system, avalanches should occur below and above the observational limits.  For this reason we probe the effect of increasing the number of vortex bundles (equivalent to decreasing the number of vortices per bundle, for fixed $\nu$) on the distribution of avalanches.

The avalanche size distributions for different values of the bundle factor $B$ are shown in the middle panels of Fig. \ref{fig:hist_N}.  The histograms are overlaid with a power law best fit.  For $B=5\times 10^{2}$ and $B=1\times 10^{5}$ there is an obvious `bump' in the glitch size distribution, whose peak occurs at the same value of $\delta\nu/\nu$ for different $B$ [$\log_{10}(\delta\nu/\nu)_{\rm{bump}} = -4$].  The bump is also present in the duration distribution (see Fig. \ref{fig:dur_B}), and there is a clear turnover in the slope of the waiting time distribution (see Fig. \ref{fig:wait_N}), which may be interpreted as a crossover from power law statistics to exponential decay.  Importantly, crossover is evidence that the system is no longer in a SOC state, as there is a preferred scale for avalanches.

In SOCS generally, there is a system-size-dependent crossover scale above which the power law distribution of glitch sizes gives way to exponential decay (see \S\ref{subsec:gen_prop}).  In a genuine critical state, the crossover scale should be an increasing function of the system size \citep{Jensen98}.  In the examples discussed by \cite{Jensen98}, the system size corresponds to the number of particles in the system (rice or sand grains, for example), not the physical extent of the system.  That the bump does not appear for small $B$ and becomes increasingly prominent for large $B$ suggests a system size effect.  On the other hand, the crossover scale, defined by the bump, does not increase with system size (\emph{i.e.} $B$) in Fig. \ref{fig:hist_N}.  Nor does system size explain why the bump is a local maximum in $p(\delta\nu/\nu)$ instead of a turnover.  In their experiments on superconductors \cite{Field} saw a crossover to steeper decay in the avalanche size distribution for large values of the magnetic field ramp rate, but not a local maximum.  If instead we interpret the system size to be the number of cells, not bundles, then the bump becomes more prominent for larger systems (e.g. with $B=1\times 10^{5}$) as in  Fig. \ref{fig:hist_N}.  Despite the bump not being present for smaller systems ($N_{\mathrm{grid}}\leq40$), there is a value of $\delta\nu/\nu$, for each $N_{\rm{grid}}$, above which $p(\delta\nu/\nu)$ falls off faster than a power law, which is the expected  behaviour beyond the crossover scale defined by \cite{Jensen98} (see Fig. \ref{fig:hist_N}).  Bumps like the ones in Fig. \ref{fig:hist_N} also arise in SOCS when one changes the morphology of the grains (e.g. rice versus sand) \citep{Frette96,Jensen98,Pruessner03}.


The violation of scale invariance is at least partly due to a mismatch in how the pinning centres (cells) and vortices (bundles) are renormalized.  The most scale invariant results, closest in spirit to a SOCS, are obtained for $B=1$ (one bundle per cell, on average).  By increasing $B$ without changing $N_{\rm{grid}}$, we are renormalising the vortices on a smaller coarse grain, whilst maintaining the coarseness of the grid, which is not a justifiable physical scenario.  \cite{Pruessner02} claimed that inappropriately chosen parameters produce a bump in the size distribution, which further suggests that $B\gg 1$ alters the fundamental statistical properties of the driven system.  The implications for the microphysics of pinning in pulsars are explored in \S\ref{sec:conc}.  

In summary, bundling in the model introduce a preferred scale into the system.  (This scale greatly exceeds the vortex quantum $\kappa$, which is too small to be resolved by a practical simulation.)  When the average number of bundles per cell is kept near one, the system exhibits scale-invariant behaviour, as demonstrated by the power laws in glitch size.  In this regime, the grid cells and bundles are well matched.  If, however, the number of cells is significantly outnumbered by the vortex bundles, the scale invariance of the system is broken.  In the absence of computational limitations, one should represent each pinning site and each vortex uniquely, eliminating the need for bundling.  On the other hand, bundling confers a distinct advantage: it circumvents the subtleties of vortex-vortex interactions and reconnections by renormalizing these effects into effective vortex-vortex forces on a coarsened grid.


\subsection{Pinning threshold}
\label{subsub:pin}

\begin{figure*}
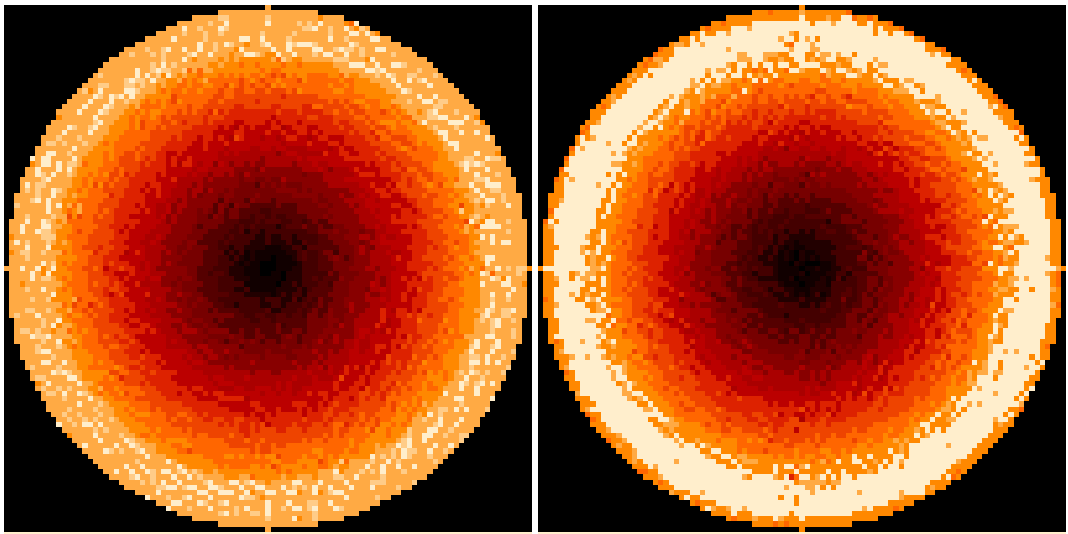

\includegraphics[width = 7cm]{fig_11a.epsi}
\includegraphics[width = 7cm]{fig_11b.epsi}
\caption{Images of the velocity imbalance $|\Delta \mathbf{v}|$ for $|\Delta \mathbf{v}_{\mathrm{pin}}|=10^{4}\rm{\,m\,s}^{-1}$ (upper) and $9\times10^{3}\rm{\,m\,s}^{-1}$ (lower).  White cells are supercritical; the darker the cell, the smaller the $\Delta \mathbf{v}$.}
\label{fig:image}
\end{figure*}

The pinning threshold $F_{\rm{pin}}$ determines the minimum lag between the superfluid and pinned vortices  necessary to unpin vortex bundles.  The pinning threshold can be reexpressed in terms of a minimum $|\Delta\mathbf{v}_{\mathrm{pin}}|$ between the pinned vortices (comoving with the crust) and the unpinned superfluid.

As $|\Delta\mathbf{v}_{\rm{pin}}|$ increases, the number of avalanches occurring within a fixed time ($10^{5}$ big time steps) decreases, from $33549$ for $|\Delta\mathbf{v}_{\rm{pin}}|=9.5\times 10^{3}$\,m\,s$^{-1}$ to $2227$ for $|\Delta\mathbf{v}_{\rm{pin}}|=10.4\times 10^{3}$\,m\,s$^{-1}$, as shown in Fig. \ref{fig:hist_nu}.  There are two reasons for this trend.  First, greater imbalance is needed in the nearest neighbour cells to overcome the pinning force and trigger an avalanche.  The greater is the required imbalance, the less probable it becomes.  Second is the issue of fine tuning.  For a given $|\Delta\mathbf{v}_{\rm{pin}}|$ there is a critical circle where the mean-field contribution to the velocity imbalance matches the pinning threshold almost exactly.  Well outside the circle, every cell is supercritical and the vortices never pin.  Near the circle, however, we find a mixture of sub- and supercritical cells because the nearest neighbour imbalance can act to reduce $|\Delta\mathbf{v}|$ at a given cell, if it opposes the mean-field contribution.  As $|\Delta\mathbf{v}_{\rm{pin}}|$ increases, the critical circle moves towards the origin, and hence the number of cells in its vicinity decreases.  Fig. \ref{fig:image} depicts images of the grid after $10^{5}$ big time steps.  The colours encode the value of $|\Delta\mathbf{v}|$ on each cell.  Supercritical cells are white; all other cells are subcritical.  The pinning threshold in the upper panel of Fig. \ref{fig:image} is $|\Delta\mathbf{v}_{\rm{pin}}|=10^{4}$\,m\,s$^{-1}$.  We observe a smattering of supercritical cells near the critical circle.  Inside the circle, all cells are subcritical.  In the lower panel of Fig. \ref{fig:image}, with $\Delta\mathbf{v}_{\mathrm{pin}}=9\times 10^{3}$m\,s$^{-1}$, we observe an annulus in which most cells are supercritical (where the vortices never pin).  Just inside the annulus, there is, like in the upper panel, a mixture of sub- and supercritical cells.  It is in these mixed regions where we claim that self-organized criticality controls the unpinning dynamics.  These results are evidence that the unpinning activity is restricted to a narrow annulus, and hence unaffected by the use of a full circle as the simulation grid, as discussed in \S\,\ref{subsec:glitch}.

A power law describes the size distribution well for all three values of $|\Delta\mathbf{v}_{\mathrm{pin}}|$ in Fig. \ref{fig:hist_nu}.  The best fit power law and its exponent are shown in each panel of Fig. \ref{fig:hist_nu}.  We find that $a$ decreases as $|\Delta\mathbf{v}_{\mathrm{pin}}|$ increases, however there is a slight upturn in both $a$ and $b$ for the largest values of $|\Delta\mathbf{v}_{\rm{pin}}|$ considered.  For larger values of $|\Delta\mathbf{v}_{\mathrm{pin}}|$, avalanches involving many cells (and hence many vortex bundles) are less likely to occur, both because the active annulus shrinks, and because it is relatively unlikely that the nearest-neighbour imbalance sends a cell supercritical.  Hence $a$ decreases because the dynamic range decreases; the area under the histogram equals the number of avalanches.

Fig. \ref{fig:velocity} demonstrates that, for a finite, square vortex array, the profile of $\Delta\mathbf{v}$ along a linear cut flattens out in the region where $r\approx R$, compared to smaller radii.  Hence if $|\Delta\mathbf{v}_{\rm{pin}}|$ approaches the maximum $\Delta\mathbf{v}$ on the grid, more cells may participate in an avalanche, in comparison to when the critical circle lies further in.  This may account for the upturn in $a$ versus $|\Delta\mathbf{v}_{\rm{pin}}|$ seen in Fig. \ref{fig:pl}.

\subsection{Pinned fraction}
\label{subsub:frac}

The pinned fraction $\epsilon$ impacts on the dynamics in two main ways:  it changes the size of each vortex bundle, and alters the relative contribution to the velocity imbalance from each term in Eq. (\ref{eq:deltav}).  Increasing the size of the vortex bundles increases the minimum avalanche size, whilst maintaining the maximum avalanche size.  Larger bundles also enhance the role of the nearest neighbour cells.  As more vortices pin and unpin, the first and third terms in Eq. (\ref{eq:deltav}) decrease with respect to the contribution from the unpinned vortices.  Given that the unpinned vortices are homogeneously distributed, they are the foremost contributors to the mean-field critical circle discussed in \S\ref{subsub:pin}.  Decreasing their relative contribution reduces the mean-field effect, emphasizing the nearest-neighbour, stochastic dynamics of the model.  The lower left panel of Fig. \ref{fig:pl} supports this argument:  the $1\sigma$ error in the fitted $a$ and $b$ values decreases with increasing $\epsilon$.  The lower left panel of Fig. \ref{fig:pl} also shows that both $a$ and $b$ increase with increasing $\epsilon$.  For $\epsilon \geq 0.0097$ the $a$ distribution flattens out, whereas the $b$ distribution continues to increase.  

For $\epsilon\leq 0.009$ no avalanches are observed; the model is badly tuned.  Moreover, for $\epsilon\gtrsim0.010$, the model slows down computationally, as a large number of cells regularly pin and unpin.  This computational limitation arises because our algorithm does not excise regions in which vortices are permanently unpinned (see the lower panel of Fig. \ref{fig:image}), instead calculating $\Delta \mathbf{v}$ unneccessarily in these regions at each small time step.  We intend to address this flaw in future work.

\subsection{Spin frequency}
\label{subsec:nu}

Two values of the spin frequency enter the model:  at birth ($\nu_{0}$), and today ($\nu$).  As a first approximation, the total number of \textit{pinned} vortices is taken to be unchanged over the lifetime of the star and hence is dictated by $\nu_{0}$ as in Eq. (\ref{eq:totv}).  The pinned portion of the superfluid does not spin down with the stellar crust, whereas the unpinned superfluid does; its circulation is dictated by $\nu$.  In the current model we hold $\nu$ and $\nu_{0}$ constant (cf. \S\ref{subsec:long_term}), with $\nu_{0}=30.25$\,Hz, and $12.73\leq\nu\leq 15.91\rm{\,Hz}$.

By decreasing $\nu$, we increase $|\Delta\mathbf{v}|$ and hence the force imbalance at every cell.  The impact is similar to decreasing $|\Delta\mathbf{v}_{\rm{pin}}|$ because the critical circle shrinks.  The two right-hand panels of Fig. \ref{fig:pl} display a similar trend for increasing $|\Delta\mathbf{v}_{\rm{pin}}|$ and $\nu$.  The main difference is that $\nu$ also determines the relative size of the first two terms in Eq. (\ref{eq:deltav}), given that the number of pinned vortices remains constant while the number of unpinned vortices decreases with $\nu$.  These effects increase the range of avalanche sizes as $\nu$ decreases as shown in Fig. \ref{fig:hist_nu}.  For large values of $\nu$, the dynamic range goes to zero, as the critical circle has moved outside the star.  For small $\nu$, a turnover emerges in the size distribution for small $\delta\nu/\nu$, which is also present in the size distributions for small values of $|\Delta \mathbf{v}_{\mathrm{pin}}|$ in Fig. \ref{fig:hist_nu}.  This may happen because the critical circle lies well within the star, so there is a sizable region outside the critical circle where vortices are permanently unpinned.  The automaton does not recognise that permanently unpinned vortices do not, in fact, participate in avalanches in real physical systems, so the output is biased towards large avalanches.

\subsection{Long-term activity}
\label{subsec:long_term}

\begin{figure*}
\includegraphics[width = 6cm,angle=90]{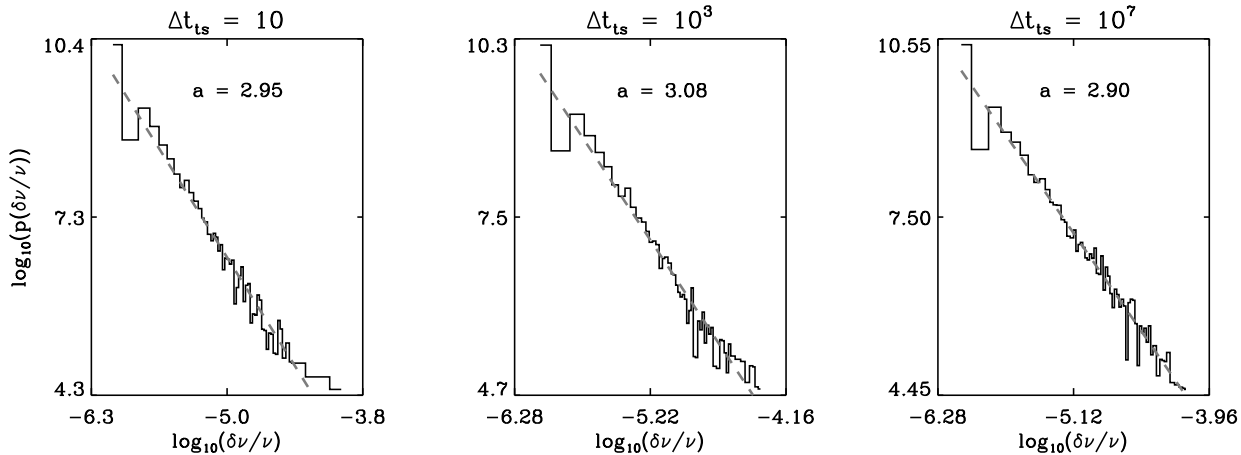}
\caption{Probability density functions $p(\delta\nu/\nu)$ of avalanche sizes $\delta\nu/\nu$, for time steps lasting $\Delta t_{\rm{ts}}=10$ (\textit{left}), $10^{5}$ (\textit{centre}) and $5\times 10^{5}$ (\textit{right}), plotted as histograms on a log-log scale, using $10^{2}$ logarithmically spaced bins.  The histograms are fitted with a power law of the form $p(\delta\nu/\nu)\propto(\delta\nu/\nu)^{-a}$ (\textit{dashed lines}).
Automaton parameters:  $N_{\mathrm{grid}}=100$, $10^{5}$ big time steps.
Physical parameters:  $|\Delta\mathbf{v}_{\rm{pin}}|=10^{4}\rm{\,m\,s}^{-1}$, $\rho_{\rm{s}}=10^{16}\rm{\,kg\,m}^{-3}$, $\nu_{0}=30.25\rm{\,Hz}$, $\nu=13.9\rm{\,Hz}$, $\epsilon=0.01$, $R=20\rm{\,km}$.}
\label{fig:hist_deltat}
\end{figure*}

\begin{figure}
\includegraphics[width = 6cm,angle=90]{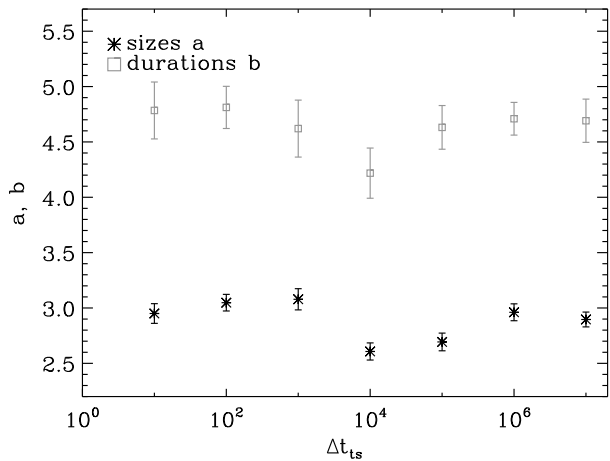}
\caption{Fitted power law exponents for the distribution of glitch sizes ($\ast$) and durations ($\square$), for time steps in the range $10\leq\Delta t_{\rm{ts}}\leq 5\times 10^{5}$.  The error bars represent the $1\sigma$ uncertainty returned by the least-squares fitting algorithm.
Automaton parameters:  $N_{\mathrm{grid}}=100$, $10^{5}$ big time steps.
Physical parameters:  $|\Delta\mathbf{v}_{\rm{pin}}|=10^{4}\rm{\,m\,s}^{-1}$, $\rho_{\rm{s}}=10^{16}\rm{\,kg\,m}^{-3}$, $\nu_{0}=30.25\rm{\,Hz}$, $\nu=13.9\rm{\,Hz}$, $\epsilon=0.01$, $R=20\rm{\,km}$.}
\label{fig:pl_deltat}
\end{figure}

To model the long-term glitch behaviour of a pulsar, we must allow $\nu$ to mimic the electromagnetic spin down of the stellar crust.  Observed spin down rates differ by several orders of magnitude; in this paper we take $\dot{\nu}=-7\times10^{-13}\,\rm{Hz\,s}^{-1}$ by way of example.  Observations of pulsar glitches span a period of approximately forty years ($\sim$1970 to present), which equates to a total spin down of $\sim 0.1$\,Hz.  In our model, each big time step must therefore span $\sim 10^{6}$\,s, in order to represent a total observing time of forty years.  Such a large time step determines the scale on which we can resolve individual avalanches.  Observationally, glitches can be resolved down to a $40$s window \citep{McCulloch90, Dodson02}.

Spin down causes the critical circle to shrink towards the axis of the star, because the number of pinned vortices is fixed, whilst the number of unpinned vortices decreases with $\nu$.  The rate of spin down thus determines how quickly the glitch behavior changes as the critical circle moves through the strata of the star.  Fig. \ref{fig:hist_deltat} shows the size distribution ($\delta\nu/\nu$) for time steps $\Delta t_{ts}=10$, $10^{5}$ and $5\times10^{5}$\,s.  Fig. \ref{fig:pl_deltat} shows the fitted power law exponents for the sizes and durations, for $10\leq\Delta t_{ts}\leq5\times10^{5}$.  Within the uncertainty of the fitting algorithm, $a$ and $b$ are the same for all values of $\Delta t_{ts}$ tested.  For the largest $\Delta t_{ts}$, the time elapsed after $10^{5}$ big time steps is $5\times10^{5}$\,s, which equates to $\Delta\nu=-0.007$\,Hz.  This moves the critical circle towards the origin by a negligible $\sim 0.007\,\%$.

The long term fate of a pulsar, with respect to its glitch behaviour, depends on the location of its pinning zones.  If pinning can occur througout the pulsar, glitches continue to occur indefinitely, as the critical circle moves inwards.  As time passes, however, the glitches become smaller, on average, as the number of cells in the vicinity of the critical circle diminishes \footnote{Interestingly, there is no guarantee that glitches emanating from different critical circles are, in fact, drawn from the same statistical distribution (e.g. value of $a$).  If so, we do not expect a pure power law in $\delta\nu/\nu$ over long times.}.  If, as claimed by \cite{Donati03}, pinning occurs at intermediate densities, no glitches occur once the critical circle shrinks inside the radius at which these densities occur.  Alternatively, if pinning occurs at several strata [high and low densities, for example \citep{Avogadro}], glitches occur when the critical circle lies in the vicinity of the pinning zones, which could mean that episodes of glitching are punctuated by `quiet' periods.

In the models of post-glitch relaxation developed by \cite{Alpar96}, the superfluid spins down when vortices move radially outward.  The outward motion is driven by the distribution of capacitive and depleted zones; outward motion is energetically favourable due to a radial bias in vortex-vortex scattering.  In our automaton, spin down is an input; it does not arise self-consistently from vortex motion.  By reducing  the number of unpinned vortices with time (e.g. \S\,\ref{subsec:long_term}) we attempt to recreate the scenario in which outward radial motion eventually makes vortices exit the pinning zones in the star.  The automaton rules do inadvertently allow for outward radial motion thanks to the rectangular nature of the grid:  azimuthal motion (i.e. in the direction of rigid body rotation) is never truly possible, as the vortices can only ever move horizontally, vertically, or diagonally.  To incorporate outward motion and hence spin down self-consistently, the automaton rules should add a radial component to the bulk superfluid flow when computing $\bf{v}_{L}$.  (This does not remove the artificial radial motion arising from the rectangular grid.  In this context, an annular grid topology is preferable.)

\section{Discussion}
\label{sec:conc}

In this paper we present a cellular automaton model for glitches in neutron stars, based on the vortex unpinning paradigm \citep{Anderson,Alpar84}.  Despite the gross idealisations in such a discrete model, the resulting statistical behaviour agrees with recent analyses of radio timing data \citep{Wang00,Wong01,Melatos07a}.  Of particular interest are the power law distributions of glitch sizes and durations produced by the model, providing further evidence that glitches are a scale invariant phenomenon.  For this reason, as observations improve in resolution and lengthen in duration, we expect to discover smaller and larger glitches than have been seen so far.  Moreover, the exponential waiting-time distribution generated by the model supports the idea that the each vortex avalanche is a statistically independent event.  The values of the power law exponents and Poisson glitching rates depend on the particular parameter set. Our fits return size and duration exponents in the ranges $2.0\leq a\leq 4.3$ and $2.2\leq b\leq 5.5$ respectively, and mean glitching rates in the range $0.0023\leq\lambda\leq0.13$.  Although we do not have a first-principles theory against which to compare our results, we claim that our cellular automaton is consistent with a SOCS.

The operation of the cellular automaton relies heavily on the assumption that vortices are inhomogeneously distributed.  \cite{Alpar96} suggested that crust cracking creates capacitive regions where many vortices pin simultaneously.  These regions do not permit a vortex current, so pinned vortices in these regions do not participate in outward vortex creep.  Crust cracking is a nonuniform process, suggesting that the regions themselves are inhomogeneously distributed.  Our model conforms qualitatively with this picture.  Of particular interest is the claim by \cite{Alpar95} that the neutron star crust is divided coarsely into depleted and capacitive regions.
If verified, this claim suggests that the level of coarse graining necessary to render our cellular automaton computationally tractable matches the level of inhomogeneity that is realistically presesnt in a neutron star.

Of course, the opposite viewpoint is also viable:  the crust forms a large, defect-free crystal, allowing the vortices to occupy a contiguous sequence of pinning positions and preventing the type of inhomogeneity represented by our initial conditions \citep{Jones98a}.  Importantly, such a scenario challenges \textit{any} model based on collective unpinning of numerous vortices, not just our cellular automaton.  \cite{DeB98} concluded that the crust is relatively free of microcrystalline structures and vacancies, with impurities, or point defects, the most likely disruptions to the crystal lattice (faults and dislocations were not considered).  Alternatively, if pinning occurs mostly at grain boundaries and cracks, as originally postulated by \cite{Anderson}, then inhomogeneity on macroscopic scales is possible.


Irrespective of the density and distribution of pinning sites in reality, the model assumes for simplicity that vortex pinning occurs throughout the pulsar.  Several recent studies have shown that pinning forces sufficient to oppose the Magnus force develop only in specific zones of the crust, e.g. regions of high and low, rather than intermediate, denisty \citep{Avogadro}.  Restricting the simulation to these optimal pinning zones obviates the need to allow for a range of pinning strengths if the zones are thin layers. A model in which pinning does not occur everywhere in the star leads to a different computational topology, e.g. an annulus, or several nested annuli.  However, we choose to be conservative in this introductory paper by not imposing a grid topology a priori, in order to see if a preferred topology emerges by itself.  In fact, it does; we find that a narrow active annulus, situated at the critical radius, participates in the avalanches.  However, it is important to experiment with other topologies in future work, because the active annulus we find in this paper may still be an artifact. If the active region lies near the rotation axis, an automaton executed in an annulus is free of the shortcomings introduced by the `permanently unpinned' regions mentioned in \S\ref{subsub:frac}.  To use the full circular grid in a meaningful way, we must include pinning of proton and neutron vortices in the core of the neutron star, together with pinning of neutron vortices in the inner crust.  In this case, the additional physics of entrainment \citep{Link96,Ruderman98,Sedrakian95,Andersson04} and induced electric currents implies a more detailed set of automaton rules, outside the scope of the model presented here.  It should also be noted that if the superfluid vortices are tangled \citep{Peralta06a,Peralta06b}, rather than forming a regular Abrikosov array, these two-dimensional annular and circular topologies are clearly too restrictive. 

The locations of pinning zones also determine the long-term fate of a pulsar as it spins down.  If pinning occurs everywhere, spin down increases the proportion of vortices with $|\Delta\mathbf{v}|>|\Delta\mathbf{v}_{\rm{pin}}|$ which are always supercritical (outside the critical circle) and do not participate in avalanche dynamics as they never pin.  If pinning is restricted to one or more annuli, the glitch behaviour of the pulsar changes significantly when the critical circle moves into and out of the pinning zones.  

Clearly, our cellular automaton does not include the response of the pulsar to glitches (ie. the spin up of the pulsar crust and the subsequent `relaxation').  Nor do we account for the finite time-scale on which the unpinned superfluid couples to the crust, governed by the Hall-Vinen-Bekarevich-Khalatnikov equations \citep{Peralta05,Peralta06a}.  We make the approximation that this time-scale is considerably less than the time-scale of our big time steps, motivated by glitch timing data (where the post-glitch recovery phase typically ends well before the next glitch; cf. Vela).

To elicit avalanches from our model, we require fine tuning in both the physical and computational parameters, such that $|\Delta\mathbf{v}_{\rm{max}}|\approx|\Delta\mathbf{v}_{\rm{pin}}|$.  This condition ensures that there are enough vortex bundles that switch between becoming sub- and supercritical as time passes.  We achieve this by choosing $\nu$, $\epsilon$ and $\Delta\mathbf{v}_{\rm{pin}}$ to place the critical circle near the surface of the star.  We emphasize that for a larger (smaller) star, we would resize the critical circle proportinally, for computational rather than physical reasons.  Fine tuning is also required in the ratio of vortex bundles to grid cells $B$.  For $B\gg 1$, the automaton output is no longer consistent with a SOCS.

In conclusion, we present an empirical cellular automaton model of pulsar glitches based on the mass vortex unpinning paradigm.  We find that for certain physical and computational parameters the model produces dynamics that are consistent with a SOCS and with radio timing data from pulsars.  There exists no general, first-principles theory of SOC , let alone of pulsar glitches, so many of our results are empirical.  In particular, ther is no way known at present to predict theoretically the size and duration distribution exponents, and the mean rate of the waiting-time distribution.  We do demonstrate that the basic physical principles governing inter-vortex interactions can produce the type of \textit{collective} behaviour necessary to explain pulsar glitches within the mass unpinning paradigm, especially the puzzle of how so many vortices can unpin in sympathy during a glitch, and why their number varies so much from glitch to glitch.

We thank Carlos Peralta for sharing his up-to-date glitch catalogue, and Stuart Wyithe for advice on statistical methods.  LW acknowledges the support of an Australian Postgraduate Award.


\bibliography{bib}
\bibliographystyle{mn2e}

\label{lastpage}







\newpage
\newpage

\end{document}